\documentclass[11pt,draftcls,onecolumn]{IEEEtran}

\usepackage{epsfig,cite}
\usepackage{amsmath,amssymb}
\usepackage{arydshln}

\newtheorem{theorem}{Theorem}

\newtheorem{example}[theorem]{Example}

\newtheorem{algorithm}[theorem]{Algorithm}

\newcommand{\norm}[1]{\lVert#1\rVert}
\newcommand{\abs}[1]{\lvert#1\rvert}
\newcommand{\field}[1]{{\mathbb{#1}}}

\newcommand{\Prob}{\mathrm{Prob}}

\newcommand{\End}{\hfill \mbox{$\bigtriangledown$}}
\newcommand{\one}{\mathbf{1}}
\newcommand{\ben}{\begin{enumerate}}
\newcommand{\een}{\end{enumerate}}
\newcommand{\ba}{\begin{array}}
\newcommand{\ea}{\end{array}}
\newcommand{\beq}{\begin{equation}}
\newcommand{\eeq}{\end{equation}}

\newcommand{\R}{\field{R}}
\newcommand{\Z}{\field{Z}}
\newcommand{\E}{\field{E}}

\newcommand{\fig}[3]{\resizebox{#1}{#2}{\includegraphics{#3}}}

\begin{document}

\title{%
 The PageRank Problem, Multi-Agent Consensus and Web Aggregation\\
{\Large A Systems and Control Viewpoint}
}

\author{%
Hideaki Ishii
and
Roberto Tempo
\thanks{%
This work was supported in part by the Ministry of Education, Culture, Sports, 
Science and Technology, Japan, under Grant-in-Aid for Scientific Research Grant 
No.~23760385, in part by the Aihara Project, the FIRST program from JSPS, initiated
by CSTP, and in part by the European Union
Seventh Framework Programme [FP7/2007-2013] under grant agreement n.257462 HYCON2 Network
of Excellence.}
\thanks{%
H.~Ishii is with the Department of Computational Intelligence 
and Systems Science, Tokyo Institute of Technology,
4259 Nagatsuta-cho, Midori-ku, Yokohama 226-8502, Japan.
(e-mail: ishii@dis.titech.ac.jp).} 
\thanks{%
R.~Tempo is with CNR-IEIIT, Politecnico di Torino, 
Corso Duca degli Abruzzi 24, 10129 Torino, Italy.
(e-mail: roberto.tempo@polito.it).}}


\maketitle



PageRank is an algorithm introduced in 1998 and used by the Google Internet search engine.
It assigns a numerical value to each element of a set of hyperlinked documents 
(that is, web pages) within the World Wide Web with the purpose of measuring 
the relative importance of the page \cite{google:13}. 
The key idea in the algorithm is to give a higher PageRank value to 
web pages which are visited often by web surfers.
On its website, Google describes PageRank as follows:
``PageRank reflects our view of the importance of web pages by considering 
more than 500 million variables and 2 billion terms. Pages that are considered
important receive a higher PageRank and are more likely to appear at 
the top of the search results."

Today PageRank is a paradigmatic problem 
of great interest in various areas, such as
information technology, bibliometrics, biology, and e-commerce, 
where objects are often ranked in order of importance. 
This article considers a distributed randomized approach based on 
techniques from the area of Markov chains 
using a graph representation consisting of nodes and links.
We also outline connections with other 
problems of current interest to the systems and control community, 
which include ranking of control journals, consensus of multi-agent 
systems, and aggregation-based techniques.

\newpage
\section*{The PageRank Problem and Its History}

The PageRank algorithm was introduced by the cofounders of Google 
in the late 1990s \cite{BriPag:98} and has been implemented on the search engine of Google
from the time of its launching. 
%
It continues to be part of the search engine at Google and is said to be
one of the 200 signals used for narrowing down the search results \cite{google:13}
(with a link to the original paper \cite{BriPag:98}).
PageRank indicates the importance of a web page determined by the hyperlink structure 
of the web. Specifically, it is determined by the number of hyperlinks pointing to 
the page as well as the importance of the pages where those hyperlinks originate.
Related ideas for ranking objects had been previously used in other contexts, 
such as sociometry \cite{Hubbell:65}, and bibliometrics \cite{PinNar:76},
and they can be traced back to 1941 
to studies of quantitative data in economics \cite{Leontief:41}.
More recently, ideas from PageRank have been used to rank other objects in order of importance, including scientific papers linked by citations \cite{CXMR:09}, authors related by co-authorship 
\cite{LiBoNede:05}, professional sport players 
\cite{Radicchi:11}, and protein in systems biology \cite{ZaBeEf:12}. 
Refer to \cite{LanMey:06} for an introduction to the PageRank problem and to
\cite{Franceschet:11} for historical notes.

The PageRank problem recently attracted the attention of the systems and control community. 
In this context, the problem was first introduced and studied in \cite{IshTem:10}, where 
a randomized decentralized approach has been proposed. Such an approach is meaningful
in view of the computational difficulties due to the size of the problem and in the
web environment where computational resources are available. 
In the abovementioned paper, the mean-square ergodic convergence properties of 
the distributed algorithms have been analyzed. 
Other convergence properties have been 
subsequently studied in \cite{ZhaChFa:13}, where almost sure convergence of the same 
decentralized scheme is demonstrated using techniques of stochastic approximation algorithms.
In \cite{NazPol:11}, a randomized algorithm based on stochastic descent is proposed 
and an explicit bound on the convergence rate is computed. 
The fluctuations of the PageRank values in the presence of fragile and uncertain 
links have been studied in \cite{IshTem_sice:09}. 
In \cite{CsaJunBlo:09}, an approach based on 
Markov decision processes is developed, and optimization and robustness viewpoints are followed 
in \cite{FABG:13} and \cite{JudPol:12}.
Motivated by the PageRank problem, the recent works \cite{Nesterov:12,Necoara:13} present
randomized distributed algorithms for solving 
huge-scale convex optimization problems with underlying sparse network structures.
Other recent references on PageRank are listed in \cite{IshTemBai:12},
where a web aggregation approach is studied, and in \cite{IshTemBai_scl:12}, where
intermittent communication in the distributed randomized computation is dealt with.
Finally, as we will see later, the PageRank problem can be viewed as finding
an eigenvector of an adjacency matrix for a web graph, 
and there have been research efforts to do this decentrally in the 
area of multi-agent systems. (See, for example, \cite{KibCom:12}.)

\section*{The Hyperlink Matrix and the Random Surfer Model}

In this section, a network consisting of web pages and hyperlinks connecting them
is described based on a graph theoretic approach.
Consider a network of $n$ web pages indexed by integers from 1 to $n$, where
$n\geq 2$ to avoid trivial situations.
This network is represented by the graph $\mathcal{G}:=(\mathcal{V},\mathcal{E})$,
where $\mathcal{V}:=\{1,2,\ldots,n\}$ is the set of vertices corresponding 
to the web page indices and $\mathcal{E}$ is the set of edges representing
the links among the pages. 
The vertex $i$ is connected to the vertex $j$ by an edge, that is,
$(i,j)\in \mathcal{E}$, if page $i$ has an outgoing link to page $j$,
or equivalently page $j$ has an incoming link from page $i$.
Since the edges have directions, the graph $\mathcal{G}$ is said to be directed.
In particular, the index set of pages linked to page $i$ is given by $\mathcal{L}_i:=\{j:\; (j,i)\in\mathcal{E}\}$ and $n_j$ is the number of outgoing links of page $j$. 

Then, the {\it hyperlink matrix} $A=(a_{ij})\in\R^{n\times n}$ is defined as
\begin{equation}
 a_{ij} 
  := \begin{cases}
      \frac{1}{n_j} & \text{if $j\in \mathcal{L}_i$},\\
      0             & \text{otherwise}.
    \end{cases}
 \label{eqn:A}
\end{equation}
The hyperlink matrix $A$ has some important properties. First, it is a nonnegative matrix, 
that is, all of its entries  $a_{ij}$ are nonnegative. This property is expressed as
$$
A\geq 0.
$$
Note that the link matrix is column-wise normalized by construction. 
However, in the real web, pages having no links to others are abundant and
are referred to {\it dangling nodes}. 
Such pages can be found, for example, in the form of PDF and image files
having no outgoing hyperlinks. These pages introduce zero columns into the link matrix. 
Then, it can be easily verified that the 
hyperlink matrix $A$ is a {\it column substochastic} matrix, that is, 
it is a nonnegative matrix $A\in \R^{n\times n}$ having  
the property that $\sum_{i=1}^n a_{ij} \le1$ for $j=1,\ldots,n$.

\begin{figure}[t]
  \centering
  \resizebox{6cm}{!}{\includegraphics{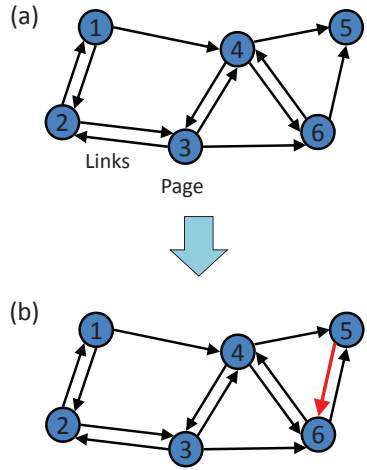}}
  \caption{The example web with six pages. (a)~Page 5 is a dangling node. 
           (b)~This is changed by addition of a back link (red) to page 6. 
            Notice that no self-loop is present.
}
\label{fig:graph}
\end{figure}

A problem related to such nodes can be explained through 
the example web in Fig.~\ref{fig:graph} (a). Here, 
the dangling node 5 behaves as a ``black hole,''
that is, the entire flow of information enters into this node and cannot escape.
Thus, the PageRank value of this page does not contribute to other pages.
To resolve this problem, the graph and consequently the matrix $A$
need to be redefined by adding artificial links for 
all dangling nodes. As a result of this modification, the columns with only zero entries 
are removed from the link matrix $A$, and this matrix $A$ becomes a {\it column stochastic} matrix, 
that is, it is a nonnegative matrix 
with the property that $\sum_{i=1}^n a_{ij}=1$ for $j=1,\ldots,n$.

The modification in the graph can be easily explained by introducing the concept of ``back button''
(that is, one returns to the previous page when visiting a dangling node).
Other methods to handle the issue have also been proposed such as replacing the zero columns
with any fixed stochastic vector and grouping all dangling nodes in one node \cite{LanMey:06,IpsSel:07}.
This is illustrated in the simple example in Fig.~\ref{fig:graph}
with a graph consisting of six web pages and twelve links. 
This example will be used throughout the article.

\begin{example}\label{ex:1}\rm
Consider the web of six pages shown in Fig.~\ref{fig:graph}\;(a). 
Note that page 5 is a dangling node since it has no outgoing links.
As previously discussed, the graph is modified by adding a link from page 5 to
page 6 (Fig.~\ref{fig:graph}\;(b)). Then, the hyperlink matrix becomes a column stochastic matrix
\begin{equation*}
 A = \begin{bmatrix}
       0           & \frac{1}{2} & 0           & 0           & 0 & 0\\
       \frac{1}{2} & 0           & \frac{1}{3} & 0           & 0 & 0\\
       0           & \frac{1}{2} & 0           & \frac{1}{3} & 0 & 0\\
       \frac{1}{2} & 0           & \frac{1}{3} & 0           & 0 & \frac{1}{2}\\
       0           & 0           & 0           & \frac{1}{3} & 0 & \frac{1}{2}\\
       0           & 0           & \frac{1}{3} & \frac{1}{3} & 1 & 0
     \end{bmatrix}.
\end{equation*}
\End
\end{example}

\vspace{.5cm}

One way to describe the PageRank problem is through 
the {\it random surfer model}: Starting from a node at random, 
the random surfer follows the hyperlink structure by picking a link at random. For example, 
if the node 3 in Fig.~\ref{fig:graph} is chosen, then the surfer goes to node 4 with probability 1/3 because 
node 3 has three outgoing links, and all links are considered equiprobable. Then,
from node 4, the surfer reaches node 6 with probability 1/3 because node 4 has three outgoing links. 
This process is then repeated. If the random surfer eventually reaches node 5, then the surfer may 
return to node 6  with probability one using the back button shown in red in 
Fig.~\ref{fig:graph}\;(b).

Mathematically, this process is described as a Markov chain 
\begin{equation}
x(k+1) = A x(k),
\label{markov}
\end{equation}
where $x(k) \in \R_+^{n}$ is a vector representing the values of the pages. 
The initial values are normalized so that $x(0)$ is 
a stochastic vector, that is,
$\sum_{i=1}^n x_i(0) = 1$. 
Since the matrix $A$ is stochastic, $x_i(k)$ is a real number 
in $[0,1]$ for all $k$ and the values are normalized so that 
$\sum_{i=1}^n x_i(k) = 1$ for all $k$. 
In fact, $x_i(k)$ is the probability of being in
node $i$ at time $k$.

In the context of the random surfer model, the PageRank values of the web pages
represent some measure on how often the web pages are visited by the random surfer. 
More precisely, the probability of visits becomes higher if web sites have links 
from {\it important} (that is, often visited) web sites and especially those
that have smaller numbers of outlinks.
In this context, 
page $i$ is said to be more important than page $j$ at time $k$ if $x_i(k) > x_j(k)$.

An interesting question is whether or not the Markov chain asymptotically converges to its 
stationary distribution for any initial nonzero values $x_i(0)$, $i\in\mathcal{V}$.
Such a distribution represents the probabilities of the random surfer visiting the
web sites, indicating the popularity of the page. 
Hence, we establish properties under which 
the vector $x(k)$ in equation (\ref{markov}) asymptotically converges as
$$
x(k) \rightarrow x^* \ \ \mbox{for} \ \ k \rightarrow \infty,
$$
where $x^*$ is referred to as the {\it PageRank value}. 
Equivalently, if convergence is achieved, then it is written 
$$
x^* = A x^*.
$$

While the previous discussion is based on a Markov chain interpretation, there is a 
simple alternative linear systems interpretation. Assuming that convergence is achieved, 
the PageRank value $x^*$ is a nonnegative unit eigenvector corresponding to the eigenvalue $1$ 
of the hyperlink matrix $A$. More formally, the PageRank problem is to compute $x^*$ 
such that
\begin{equation}
x^* = A x^*,
\end{equation}
where $x^*\in [0,1]^n$ and $\sum_{i=1}^n x^*_i = 1$. In general, for existence and uniqueness 
of the eigenvector $x^*$,  it is sufficient that the web as a graph is strongly connected
\cite{HorJoh:85}, that is, for any two vertices $i,j\in {\mathcal V}$, there is a sequence of
edges which connects $i$ to $j$. In other words, from every page, 
every other page can be reached through a connected path. 
In terms of the link matrix $A$, strong connectivity 
of the graph is equivalent to $A$ being an irreducible matrix; 
see ``Definitions and Properties of Stochastic Matrices.''
For stochastic matrices, there exists at least one eigenvalue equal to 1. 
However, it is easy to verify that the web is in general not strongly connected
since there are many pages in the web that cannot be visited from another page
by simply following links. Therefore, the problem should be 
modified as described in the next section.

\section*{The Teleportation Model}

The idea of the teleportation model is that the random surfer, after a while, becomes bored and 
he/she stops following the hyperlink structure as previously described. That is, at some time instant, 
the random surfer ``jumps" to another page not directly connected to the one currently being visited. 
The new page can be in fact completely unrelated 
topically or geographically to
the previous page. All $n$ pages have the same probability $1/n$ to be reached 
by the jump. For example, in Fig.~\ref{fig:graph} the random surfer may be teleported from 
node 5 to node 1, and the probability to reach node 1 is 1/6 because all nodes are equiprobable.

Mathematically, the teleportation model is represented as a convex combination of two matrices. 
Let $m$ be a parameter such that $m\in(0,1)$, and let the modified link matrix $M\in \R^{n\times n}$ 
be defined by
\begin{equation}
M := (1-m)A + \frac{m}{n} \one \one^T,
\label{eqn:M}
\end{equation}
where $\one :=[1\,\cdots\,1]^T\in\R^n$ is a vector whose entries are all equal to $1$ 
and thus $\one \one^T \in \R^{n\times n}$ is a rank one matrix with all entries being $1$. 
Equation \eqref{eqn:M} is often referred to as the {\it PageRank equation}. The matrix $M$ therefore 
is the convex combination of the original hyperlink matrix $A$ and the matrix 
$(1/n) \one \one^T$. The latter matrix indicates that the probability of the jump is equal for all web pages, 
that is, every page can be reached by teleportation with uniform probability equal to $1/n$.
In the original algorithm \cite{BriPag:98}, 
a typical value was indicated to be $m=0.15$. 
In this article, following classical literature \cite{LanMey:06},
the same value is used
(some comments regarding why this value is used are provided at the end of this section).
Notice that $M$ is a stochastic matrix with all positive entries
because it is a convex combination of 
two stochastic matrices by $m \in (0,1)$ and $(1/n) \one \one^T$ is a positive matrix.

By Perron's theorem \cite{HorJoh:85}, 
this matrix is a primitive matrix, 
which is an irreducible matrix having only one eigenvalue of maximum modulus.
In particular, the eigenvalue 1 is of multiplicity~1 and 
is the unique eigenvalue with the maximum absolute value. Furthermore,
the corresponding eigenvector is positive; 
for more details, see ``Stochastic Matrices and Perron's Theorem.''
Therefore, the PageRank vector $x^*$ is redefined by using $M$ in place of $A$ so that
\begin{equation}
 x^* = M x^*
   ~\Leftrightarrow~ 
   x^* = (1-m)A x^* + \frac{m}{n}\one,
\label{eqn:prvec}
\end{equation}
where $x^*\in (0,1)^n$ and $\sum_{i=1}^n x^*_i = 1$. 
It now follows that the asymptotic probability $x^*$ of being in a state is independent of the 
initial state $x(0)$. Intuitively, a positive stochastic matrix represents 
a Markov chain that is not periodic and have no sinks states.
This means that the application of the stochastic matrix to a probability 
distribution would redistribute the probability mass of the original distribution 
while preserving its total mass. If this process is applied repeatedly the distribution converges 
to a stationary distribution for the Markov chain.

Example~\ref{ex:1} is now revisited to show the computation 
of the matrix $M$ and of PageRank $x^*$.

\begin{example}\label{ex:2}\rm
The matrix $M$ can be computed from equation (\ref{eqn:M}) as
\begin{equation*}
 M = \begin{bmatrix}
       0.025 &   0.450  & 0.025 & 0.025 & 0.025 & 0.025\\
       0.450  &   0.025 & 0.308 & 0.025 & 0.025 & 0.025\\
       0.025 &   0.450  & 0.025 & 0.308 & 0.025 & 0.025\\
       0.450  &   0.025 & 0.308 & 0.025 & 0.025 & 0.450\\
       0.025 &   0.025 & 0.025 & 0.308 & 0.025 & 0.450\\
       0.025 &   0.025 & 0.308 & 0.308 & 0.875 & 0.025
     \end{bmatrix}.
\label{eqn:ex:origM}
\end{equation*}
Observe that $M$ is not a sparse matrix and its diagonal entries are non-zero; 
see further comments later in this section.
The PageRank vector $x^*$ in \eqref{eqn:prvec} is
\begin{equation}
  x^* = \begin{bmatrix}
           0.0614 & 0.0857 & 0.122 & 0.214 & 0.214 & 0.302
        \end{bmatrix}^T.
\label{eqn:ex:x_ast}        
\end{equation}
Notice that pages~4 and 6 have the largest number of incoming links,
resulting in large PageRank values.
Page~6 is more advantageous because the pages contributing to its value
via links, that is, pages~3, 4, and 5, 
have larger values than those having links to page~4. 
Page~1 has the smallest number
of incoming links and obviously the lowest ranking in this web. 
It is interesting that pages~4 and 5 share the same value. 
\End
\end{example}

\vspace{.5cm}

\centerline{
\fbox{\parbox{16cm}{
{\Large Definitions and Properties of Stochastic Matrices}\\ \\
A matrix $X\in\R^{n\times n}$ in which all entries are nonnegative real numbers 
is said to be nonnegative, and it is denoted as $X \ge 0$; a matrix whose entries are 
positive real numbers is called positive, denoted as $X >0$. 
A stochastic matrix (also termed probability matrix or Markov matrix) 
is a matrix used to describe the transitions of a Markov chain. Each of its entries 
is a nonnegative real number representing a transition probability. \\ \\
A column stochastic matrix is a matrix with each column summing to one, 
and a row stochastic matrix is a matrix with each row summing to one. 
A doubly stochastic matrix has the property that each row and column sum to one. 
A stochastic vector (also called probability vector) is a vector whose elements 
are nonnegative real numbers which sum to one.\\		\\
A matrix $X\in\R^{n\times n}$ is said to be reducible if either (i)~$n=1$ and $X=0$ or (ii)~$n\geq 2$ and there exist
a permutation matrix $P\in\R^{n\times n}$ and an integer $r$
with $1\leq r \leq n-1$ such that
\[
P^T X P
  = \begin{bmatrix}
      B & C\\
      0 & D
    \end{bmatrix},
\]
where $B\in\R^{r\times r}$, $C\in\R^{r\times (n-r)}$, and 
$D\in\R^{(n-r)\times(n-r)}$. An irreducible matrix is a matrix that is not reducible.\\ \\
A nonnegative matrix is said to be primitive 
if it is irreducible and has only one eigenvalue of maximum modulus.
In the terminology of Markov chains, these conditions correspond 
to a chain being irreducible and aperiodic.
} 
}
}

\vspace{.5cm}

\centerline{
\fbox{\parbox{16cm}{
{\Large Stochastic Matrices and Perron's Theorem}\\ \\
Let $M \in \R^{n\times n}$ be a nonnegative stochastic matrix. Then, 1 is an eigenvalue of $M$ and 
there is a nonnegative eigenvector $x \ge 0, x \not =0$, such that $Ax=x$.
In this case, the eigenvector is not necessarily uniquely determined.\\ \\
Let $M \in \R^{n\times n}$ be a positive stochastic matrix. Then, the following statements hold
based on Perron's theorem \cite{HorJoh:85}:\\
1. The eigenvalue $1$ of $M$ is a simple eigenvalue such that any other eigenvalue  $\lambda_i$ 
(possibly complex) is strictly smaller than $1$ in absolute value, $|\lambda_i| < 1$. The spectral 
radius $\rho(M)$ is equal to $1$.\\
2. There exists an eigenvector $x^*$ of $M$ with eigenvalue $1$ such that all components of $x^*$ 
are positive $x^* = M x^*$, $x^*_i > 0$ for $1 \le i \le n$. \\ 
3. $M$ is irreducible and
the corresponding graph ${\mathcal G}=(\mathcal{V},\mathcal{E})$ is 
strongly connected, that is, for any two vertices $i,j\in\mathcal{V}$, there 
exists a sequence of edges which connects $i$ to $j$.\\ \\
A stationary probability vector $x^*$ is the eigenvector of the positive stochastic matrix $M$ 
associated with eigenvalue $1$; it is a vector that does not change under application of the transition matrix. 
Perron's theorem ensures the following:\\
1. Such a vector $x^*$ exists and it is unique.\\
2. The eigenvalue with the largest absolute value $|\lambda_i|$ is always $1$. \\ \\
The vector $x^*$ can be asymptotically computed by means of the power method
\begin{equation}
x(k+1) = M x(k)
\label{markov_M}
\end{equation}
for any $x(0)$ which is a stochastic vector. 
Therefore, the following limit is obtained:
\begin{equation}
x(k) \rightarrow x^*
\label{eqn:conv}
\end{equation}
for $k \rightarrow \infty$. 
} 
}
}

\vspace{.5cm}

As previously discussed, due to the large dimension of the link matrix $M$ (currently of 
the order of $10^{10} \times 10^{10}$), the computation of the PageRank values is very difficult. 
The solution that has been employed in practice 
is based on the power method, which is simply the Markov chain iteration (\ref{markov}) with $A=M$,
or (\ref{markov_M}).
It appears that this computation is performed at Google once a month and it takes one week, 
even though the power method requires only 50--100 iterations
to converge to a reasonable approximation \cite{LanMey:06}.  

The value vector $x^*$ is computed through the recursion 
\begin{align}
x(k+1) 
  &= M x(k) \nonumber\\
  &= (1-m)A x(k) + \frac{m}{n}\one,
    \label{eqn:xM0} 
\end{align}
where the initial condition $x(0)\in \R^n$ is a probability vector.  
The equality in \eqref{eqn:xM0} follows immediately from the fact 
$\one^Tx(k)=1$.
For implementation, it is much more convenient to use the form 
on the right-hand side of 
(\ref{eqn:xM0}) involving the matrix $A$ and not the matrix $M$ because the matrix 
$A$ is sparse, while the matrix $M$ is not. Also notice that $M$ has non-zero elements 
in the diagonal, and this means that self-loops are artificially introduced in 
the teleportation model, which are in fact absent in the matrix $A$.

The convergence rate of the power method is now discussed.
Denoting by $\lambda_1(M)$ and $\lambda_2(M)$, respectively,  
the largest and the second largest eigenvalue of $M$ 
in magnitude, the asymptotic rate of convergence of this method is exponential and
depends on the ratio $|\lambda_2(M)/\lambda_1(M)|$.
Since $M$ is a positive stochastic matrix, 
it holds that $\lambda_1(M)=1$ and $|\lambda_2(M)|<1$.
Furthermore, it can be easily shown that 
the structure of the link matrix $M$ leads us to the bound 
\begin{equation*}
|\lambda_2(M)| \leq 1-m = 0.85.
\end{equation*}
Therefore, after $50$ iterations the error level is below 
$0.85^{50} \approx 2.95\times 10^{-4}$, and 
after $100$ iterations, it becomes $0.85^{100} \approx 10^{-8}$.
Clearly, larger values of $m$ imply faster convergence. 
However, when $m$ is large, the emphasis on the link matrix $A$ 
and hence differences among the pages are reduced in the PageRank values. 
On the other hand, 
by performing a sensitivity analysis with respect to the parameter $m$, 
it follows that
$$
\left|\frac{{\mathrm d}}{{\mathrm d} m} x_i^*(m)\right| \le \frac{1}{m} \le 6.66
$$
for $i=1,2, \ldots,n$. A deeper analysis \cite{LanMey:06} shows that if $|\lambda_2(M)|$ 
is close to $|\lambda_1(M)|$, then 
the values in $x^*(m)$ become sensitive and may change even for small 
variations in $m$.
The conclusion is that $m=0.15$ is a reasonable compromise, 
and this is probably the reason why it is used at Google.

\section*{Distributed Randomized Algorithms for PageRank Computation}

This section studies a sequential distributed randomized approach of {\it gossip-type} which, 
at each step of the sequence, uses only the outgoing links connecting a specific web page to 
the neighboring nodes to compute the PageRank vector $x^*$ \cite{IshTem:10}. That is, in contrast with  
the centralized approach (\ref{eqn:xM0}) based on the power iteration, only local information 
involving a specific web page (randomly selected) is utilized to update the PageRank value. 
Difficulties in computing PageRank have motivated various studies on efficient algorithms,
but decentralized schemes over networks \cite{BerTsi:89} 
are natural especially in the context of web data \cite{ZhaChFa:13,Nesterov:12,Necoara:13}. 

Consider the web with $n$ pages represented by the directed graph $\mathcal{G}=(\mathcal{V},\mathcal{E})$. 
The randomized scheme is described as follows:
At time $k$, page $i$ is randomly selected and its PageRank value is transmitted by means of outgoing 
links to the pages that are directly linked to it, while other pages not directly connected to 
page $i$ are not involved in the transmission.
More precisely, we introduce a random process $\theta(k)\in\mathcal{V}$, $k\in\Z_+$, and, if 
at time $k$, $\theta(k)$ is equal to $i$, then page $i$ initiates the broadcasting process 
involving only the neighboring pages connected by outgoing links. All pages involved in this 
algorithm renew their values in a random fashion based on the latest available information.  

Specifically, $\theta(k)$ is assumed to be an independent and identically distributed (i.i.d.)
random process and its probability distribution is given by
\begin{equation}
  \Prob\{\theta(k)=i\} = \frac{1}{n},~~k\in\Z_+.
\label{eqn:theta1}
\end{equation}
In other words, at time $k$, the page starting the transmission process is selected with equal probability.  In principle, 
this scheme may be implemented without requiring 
a decision maker or any fixed order among the pages. Extensions of this scheme are studied in \cite{IshTem:10}
where multiple updates of web pages are considered. 
In this case, each page decides to update or not 
in an i.i.d.\ fashion under a given probability, independently of other pages. Furthermore, in \cite{IshTemBaiDab:09}
other more sophisticated schemes are presented.


Instead of the centralized scheme (\ref{eqn:xM0}), which involves the full matrix $A$, 
consider a randomized distributed update scheme for PageRank computation of the form
\begin{equation}
  x(k+1) 
    = (1-\hat{m}) A_{\theta(k)} x(k) 
        + \frac{\hat{m}}{n}\one,
  \label{eqn:xMi1}
\end{equation}
where the initial condition $x(0)$ is a probability vector, $\theta(k)\in\{1,\ldots,n\}$ is 
the random process defined in (\ref{eqn:theta1}), $\hat{m}\in(0,1)$ is a design parameter 
which replaces the parameter $m=0.15$ used in (\ref{eqn:xM0}), and 
$A_i$, $i=1,\ldots,n$, are the {\it distributed hyperlink matrices} of gossip-type subsequently 
defined in ``Distributed Link Matrices and Their Average.''

The objective of this distributed update scheme 
is to design the distributed hyperlink matrices $A_i$ and the parameter $\hat{m}$
so that the PageRank values are computed
through the time average of the state $x(k)$. To this end, let $y(k)$ be the time average of 
the sample path $x(0),\ldots,x(k)$ defined as
\begin{equation}
y(k) 
:= \frac{1}{k+1}\sum_{\ell=0}^{k} x(\ell).
\label{eqn:yk}
\end{equation}
For the distributed update scheme, the objective is to compute the PageRank value $x^*$ using 
the {\it time average} $y(k)$, also called the Ces\`aro average or the Polyak average in some contexts.
For each initial state $x(0)$ that is a probability vector, 
$y(k)$ is said to converge to $x^*$ in the mean-square error (MSE) sense if
\begin{equation*}
\E\left[
   \bigl\|
      y(k) - x^*
   \bigr\|^2
  \right] \rightarrow 0,~~k\rightarrow\infty,
 \label{eqn:thm:erg}
\end{equation*}
where the expectation $\E[\,\cdot\,]$ is taken with respect to the random process
$\theta(k)\in\mathcal{V}$, $k\in\Z_+$, defined in (\ref{eqn:theta1}).
This type of convergence is called ergodicity for random processes \cite{PapPil:02}.

\subsection*{Distributed Link Matrices and Their Average}

Here, the gossip-type distributed link matrices are introduced.
Based on the definition (\ref{eqn:A}) of 
the hyperlink matrix, recall that the $i$th column of the matrix $A$ represents the outgoing links 
of page $i$. Therefore, the idea of the distributed randomized algorithm is that the matrix $A_i$ 
uses only the column $i$ of the matrix $A$, and the remaining columns of $A_i$ are constructed 
so that the matrix $A_i$ is a stochastic matrix. 
This is a key feature of the method, distinguishing it from others as \cite{Nesterov:12,Necoara:13}. 
More precisely, the {\it distributed link matrices}
$A_i\in\R^{n\times n}, i=1,2, \ldots,n$, are defined as follows: \\
(i) The $i$th column of $A_i$ coincides with the $i$th column of $A$. \\
(ii) The $j$th diagonal entry of $A_i$ is equal to $1$ for 
$j=1,\ldots,n$, $j\neq i$. \\
(iii) All of the remaining entries $a_{ij}$ are zero.\\
By construction, it follows immediately that the distributed matrices $A_i$ are column stochastic. 
The next example shows the construction of the distributed link matrices.

\begin{example}\label{ex:3}\rm
In the six-page web of Example~\ref{ex:1}, 
the distributed link matrices $A_i$, $i=1,\ldots,6$, can be obtained as
\begin{align*}
A_1
  &= \begin{bmatrix}
       0           & 0 & 0           & 0 & 0 & 0\\
       \frac{1}{2} & 1 & 0           & 0 & 0 & 0\\
       0           & 0           & 1           & 0 & 0 & 0\\
       \frac{1}{2} & 0           & 0           & 1 & 0 & 0\\
       0           & 0           & 0           & 0 & 1 & 0\\
       0           & 0           & 0           & 0 & 0 & 1
     \end{bmatrix},~~
 A_2
  = \begin{bmatrix}
       1 &  \frac{1}{2}& 0           & 0 & 0 & 0\\
       0 & 0           &   0& 0 & 0 & 0\\
       0           &  \frac{1}{2}& 1 & 0 & 0 & 0\\
       0           & 0           & 0           & 1 & 0 & 0\\
       0           & 0           & 0           & 0 & 1 & 0\\
       0           & 0           & 0           & 0 & 0 & 1
     \end{bmatrix},~~
 A_3
  = \begin{bmatrix}
       1 & 0 & 0           & 0           & 0 & 0\\
       0 & 1           & \frac{1}{3} & 0           & 0 & 0\\
       0 & 0 & 0           & 0 & 0 & 0\\
       0 & 0           & \frac{1}{3} & 1 & 0 & 0\\
       0 & 0           & 0           & 0           & 1 & 0\\
       0 & 0           & \frac{1}{3} & 0           & 0 & 1
     \end{bmatrix},
\end{align*}
\begin{align*}
A_4
  &= \begin{bmatrix}
       1           & 0 & 0           & 0 & 0 & 0\\
       0 & 1 & 0           & 0 & 0 & 0\\
       0           &   0         & 1           &  \frac{1}{3} & 0 & 0\\
       0 & 0           & 0           & 0 & 0 & 0\\
       0           & 0           & 0           &  \frac{1}{3} & 1 & 0\\
       0           & 0           & 0           &  \frac{1}{3} & 0 & 1
     \end{bmatrix},~~
 A_5
   = \begin{bmatrix}
       1 &  0& 0           & 0 & 0 & 0\\
       0 & 1           &   0& 0 & 0 & 0\\
       0           &  0& 1 & 0 & 0 & 0\\
       0           & 0           & 0           &  1& 0 & 0 \\
       0           & 0           & 0           & 0 & 0 &  0\\
       0           & 0           & 0           & 0 & 1 &  1
     \end{bmatrix},~~
 A_6
   = \begin{bmatrix}
       1 & 0 & 0           & 0           & 0 & 0\\
       0 & 1           & 0 & 0           & 0 & 0\\
       0 & 0 & 1           & 0 & 0 & 0\\
       0 & 0           & 0 & 1 & 0 & \frac{1}{2}\\
       0 & 0           & 0           & 0           & 1 & \frac{1}{2}\\
       0 & 0           & 0 & 0           & 0 & 0
     \end{bmatrix}.
\end{align*}
\End
\end{example}

\medskip
The number of nonzero entries in the matrix $A_i$ is no more than $2n-1$, 
and $n-1$ diagonal entries are equal to 1. On the other hand, the centralized matrix $A$ 
has at most $n^2$ nonzero entries. The sparsity of the matrices $A_i$ is useful 
from the computational and implementation viewpoint.
Furthermore, the protocol is gossip-type because 
the link matrix $A_i$, using only the $i$th column of the centralized matrix $A$,
is chosen randomly at each time instant. 



\subsection*{Mean-Square Error Convergence of the Distributed Update Scheme}

We now discuss 
the convergence properties of the randomized distributed scheme \eqref{eqn:xMi1}, 
where the parameter $\hat{m}$ is chosen as
\begin{equation*}
   \hat{m} := \frac{2m}{n - m(n-2)}.
\end{equation*}
For the value $m=0.15$ used in \eqref{eqn:xMi1},
it holds that $\hat{m} = 0.3/(0.85n+0.3)$. 

It has been shown \cite{IshTem:10} that the time average of the randomized 
update scheme defined in \eqref{eqn:xMi1} and \eqref{eqn:yk} to compute PageRank 
converges to the value vector $x^*$ in the mean-square error sense. 
More precisely, for any stochastic vector $x(0)$, it holds
$$
\E\bigl[
   \bigl\|
      y(k) - x^*
   \bigr\|^2
  \bigr] \rightarrow 0,~~k\rightarrow\infty.
$$
The time average is necessary and, without averaging the values, 
$x(k)$ oscillates and does not converge to the stationary value $x^*$.

Several remarks are in order. 
In the distributed computation discussed here,
it is required that pages communicate with each other and 
then make computation for the updates in the values. 
More in detail, for each page, the values of the 
pages that are directly connected to it by outgoing links need to be
sent. The link matrices $A_i$ involved in the update scheme \eqref{eqn:xMi1}
are sparse. Thus, at time $k$, communication 
is required only among the pages corresponding to the nonzero entries
in $A_{\theta(k)}$. 
Then, for each page, weighted addition of its own value, the values 
just received, and the constant term $\hat{m}/n$ is performed.
Consequently, the amount of computation required 
for each page is very limited at any time.

Implementation issues, such as how web pages 
make local computations, are outside the scope of this article. 
It is, however, clear that in practice, servers hosting web pages should be making the
local computations along with the communications, and not the individual pages.
Regulations may also be necessary so that page owners cooperate with the 
search engine and the PageRank values computed by them 
can be trusted. In the consensus literature, related issues 
involving cheating have been studied. 
An example is the Byzantine agreement problem,
where there are malicious agents who send
confusing information so that 
consensus cannot be achieved; see \cite{TemIsh:07} for a discussion
on a Las Vegas approach and \cite{TemCalDab_book} 
for a detailed analysis of randomized algorithms.
Another issue concerning reliability of PageRank is link spam, that is,
artificial manipulation of the rank by adding spurious links.  
There are methods \cite{LanMey:06,AndBorHop:07}
to detect link spamming.

\vspace{.5cm}

\centerline{
\fbox{\parbox{16cm}{
{\Large Distributed Link Matrices and Convergence of the Randomized Algorithm}\\ \\
Let $\theta(k)$ be an independent and identically distributed (i.i.d.) random process with probability distribution given by
\begin{equation*}
  \Prob\{\theta(k)=i\} = \frac{1}{n},~~k\in\Z_+.
\end{equation*}
Then, the average matrix is given by
$$
\overline{A}:= \E[A_{\theta(k)}] = \frac{1}{n} \sum_{i=1}^n A_i,
$$
where the expectation $\E[\,\cdot\,]$ is taken with respect to the random process 
$\theta(k)$ defined in (\ref{eqn:theta1}). 
Then, the following properties hold:
\begin{enumerate}
\item[(i)]  $\overline{A} = \frac{2}{n} A + \frac{n-2}{n} I$, where $I$ is the identity matrix.
\item[(ii)] There exists a vector $z_0\in\R_+^n$ 
which is
an eigenvector corresponding to the eigenvalue 1 for both 
matrices $A$ and $\overline{A}$.
\end{enumerate}
Therefore, even though $A$ and $\overline{A}$ are  
completely different matrices, they share a common eigenvector for the eigenvalue 1,
which corresponds to the PageRank vector.
In showing the properties above, it is important that no self-loop is allowed in 
the web graph. \\ \\
Corresponding to the distributed update scheme in \eqref{eqn:xMi1}, 
the link matrices are given by 
$$
M_{\theta(k)} := (1-\hat{m})A_{\theta(k)} + \frac{\hat{m}}{n}\one\one^T,
$$
where the parameter $\hat{m}$ is
$$
\hat{m}:= \frac{2m}{n - m(n-2)}.
$$
Then, the average matrix of $M_{\theta(k)}$ can be expressed as
$$
\overline{M}= \E[M_{\theta(k)}] = \frac{1}{n} \sum_{i=1}^n M_i.
$$
This average matrix satisfies the following properties:
\begin{enumerate}
\item[(i)] 
$\overline{M} :=\frac{2}{n - m(n-2)} M + \left(1 - \frac{2}{n - m(n-2)}
\right)I$.
\item[(ii)]
The eigenvalue 1 is simple and is the unique eigenvalue of maximum modulus.
\item[(iii)]
The corresponding eigenvector is the PageRank value $x^*$ in \eqref{eqn:prvec}.
\\
\end{enumerate}
The randomized update scheme defined in 
\eqref{eqn:xMi1} and \eqref{eqn:yk} to compute PageRank 
has the following mean-square error (MSE) property for any $x(0)$ which is a stochastic vector:
$$
\E\bigl[
   \bigl\|
      y(k) - x^*
   \bigr\|^2
  \bigr] \rightarrow 0,~~k\rightarrow\infty.
$$
} 
}
}

The next section discusses how ideas from PageRank have been
successfully used in the context of bibliometrics.

\section*{Ranking (Control) Journals}

The {\it Impact Factor} (IF) is frequently used
for ranking journals in order of importance.
The IF for the year 2012, which is the most recently available, is defined as follows:
$$
\mbox{IF 2012} 
   := \frac{\mbox{\# citations in 2012 of articles published in the years 2010--2011}}{%
                 \mbox{\# articles published in the years 2010--2011}}.
$$
In this criterion, there is a census period (2012) of one year and a window 
period (2010--2011) of two years.  
More precisely, this is the 2-year IF. Another criterion, the 5-year IF, 
is defined accordingly, but is not frequently used (since it was introduced more
recently).
The IF is a ``flat criterion" which does not take into account 
where the citations come from, that is, if the citations arrive 
from very prestigious journals or in fact if they are positive or negative citations.

On the other hand, different indices may be introduced using ideas from PageRank. 
The random walk of a journal reader is similar to the walk described by the 
random surfer moving continuously on the web. Therefore, 
journals and citations can be represented as a network with 
nodes (journals) and links (citations). 
Such a situation is described in \cite{WesBerBer:10}: 
\begin{quote}
``Imagine that a researcher is to spend all eternity in the library randomly 
following citations within scientific periodicals. The researcher begins by 
picking a random journal in the library. From this volume a random citation 
is selected. The researcher then walks over to the journal referenced by this 
citation. The probability of doing so corresponds to the fraction of
citations linking the referenced journal to the referring one, 
conditioned on the fact that the researcher starts from the referring
journal. From this new volume the researcher now selects another 
random citation and proceeds to that journal. 
This process is repeated ad infinitum." 
\end{quote}
An interesting question can be posed as,
What is the probability that a journal is cited? To address this question, a different criterion for ranking journals, 
called the {\it Eigenfactor Score} 
(EF), has been first introduced in \cite{Ber:07}. This criterion is one of the official metrics 
in the Science and Social Science Journal Citation Reports
published
by Thomson Reuters for ranking journals; see \cite{Franceschet:13} for details.

The details of EF are now described. First an adjacent matrix 
$D=(d_{ij})\in\R^{n\times n}$ is defined as follows: 
\begin{equation*}
 d_{ij} 
  := \begin{cases}
      \mbox{\# citations in 2012 from journal $j$ to articles published in journal $i$ in 2007--2011} & \text{if $i \not = j$},\\
      0             & \text{if $i=j$},
    \end{cases}
\end{equation*}
where $n$ represents the total number of journals under consideration, which is currently 
over 8,400.


In this case, the window period is five years. The adjacent matrix $D$ is then 
suitably normalized to obtain the {\it cross-citation} matrix $A=(a_{ij})\in\R^{n\times n}$ 
as follows:
$$
a_{ij}:= \frac{d_{ij}}{\sum_{k=1}^n d_{kj}}, i, j=1,\ldots,n,
$$
where $a_{ij} =0$ if $a_{ij} =0/0$. Clearly, the fact that the diagonal entries of the matrices 
$D$ and $A$ are set to zero means that self-citations are omitted. Furthermore, the normalization 
used to obtain the matrix $A$ implies that this matrix is column substochastic.

In the unlikely event that there are journals that do not cite any other journal, 
some columns in the cross-citation matrix 
are identically 
equal to zero making the matrix substochastic instead of stochastic,
similarly to the situation in the web.
To resolve the issue, a trick similar to the ``back button" 
previously introduced can be useful.
To this end, let the {\it article vector} be given by
$v:= [v_1 \,\cdots\, v_n]^T\in\R^n$, where
$$
v_i := \frac{\mbox{\# articles published in journal $i$ in  2007--2011}}{%
               \mbox{\# articles published by all journals in 2007--2011}}.
$$
That is, $v_i$ represents the fraction of all published articles coming from journal $i$ during 
the window period 2007--2011. Clearly, $v$ is a stochastic vector. To resolve the substochasticity problem, 
the cross-citation matrix $A$ is redefined replacing the columns having entries equal to zero with 
the article vector. More precisely, a new matrix $\tilde A$ is introduced as 
$$
A = \left[ 
\begin{array}{cccccc}
a_{11} & a_{12} & \cdots & 0 & \cdots & a_{1n} \\
a_{21} & a_{22} & \cdots & 0 & \cdots & a_{2n} \\
\vdots & \vdots & \ddots & 0 & \ddots & \vdots \\
a_{n1} & a_{n2} & \cdots & 0 & \cdots & a_{nn}
\end{array}
\right]~~
\Rightarrow~~ \tilde 
A := \left[ 
\begin{array}{cccccc}
a_{11} & a_{12} & \cdots & v_1 & \cdots & a_{1n} \\
a_{21} & a_{22} & \cdots & v_2 & \cdots & a_{2n} \\
\vdots & \vdots & \ddots & \vdots & \ddots & \vdots \\
a_{n1} & a_{n2} & \cdots & v_n & \cdots & a_{nn}
\end{array}
\right].
$$
The matrix $\tilde A$ is a stochastic matrix, and therefore the 
eigenvector corresponding to the largest eigenvalue, which is equal 
to one, exists. However, to guarantee its uniqueness, 
a teleportation model similar to that 
previously described in (\ref{eqn:M}) needs to be introduced.
In this case, consider the {\it Eigenfactor Score 
equation}
$$
M := (1-m) \tilde A + m v \mathbf{1}^T,
$$
where the convex combination parameter $m$ is equal to $0.15$. 
This equation has the same form 
as the PageRank equation (\ref{eqn:M}), but the matrix $\one \one^T$ is replaced with the rank-one matrix $v \one^T$. 
The properties previously discussed for the PageRank equation hold because the matrix $M$ is a positive 
stochastic matrix. In particular, the eigenvector $x^*$ corresponding to the largest eigenvalue, 
which is equal to 1, is unique. That is,
$$
x^* = M x^*.
$$
The value $x^*$ is called the {\it journal influence vector}, 
which provides suitable weights on the citation values. 

The interpretation from the point of view of Markov chains is that the value
$x^*$ represents the steady state fraction of time spent vising each journal 
represented in the cross-citation matrix $A$. The Eigenfactor 
Score EF is an $n$-dimensional vector whose $i$th 
component is defined as the percentage of the total weighted citations that journal $i$ receives 
from all 8,400 journals. That is, we write
$$
\mbox{EF} := 100 \ \frac{A x^*}{\sum_{i=1}^n (A x^*)_i}.
$$

A related index, used less frequently, is the {\it Article Influence} AI, 
which is a measure of the citation influence 
of the journal for each article. Formally, 
the $i$th entry of the vector AI is defined by
$$
\mbox{AI}_i := 0.01 \frac{\mbox{EF}_i}{v_i}, i=1,\ldots,n.
$$

In order to preserve sparsity of the matrix $A$, from the computation point of view, we notice that 
EF can be obtained without explicitly using the matrix $M$. 
That is, the journal influence 
vector iteration is written by means of the power method 
\begin{equation*}
x(k+1) = (1-m) \tilde A x(k) + m v.
\end{equation*}

To conclude this section, a table is shown, summarizing 
the 2012 ranking of 10 mainstream control journals according to the IF and the EF.


\begin{table}[htb]
\caption{2012 Impact Factor (IF) and 2012 Eigenfactor Score 
(EF)}
\label{IFandEI}
\begin{tabular}{|c|c|c|c|c|} \hline
IF & Journal & Ranking & Journal & EF\\ \hline
2.919 & Automatica & 1 & IEEE Transactions Automatic Control & 0.04492\\
2.718 &IEEE Transactions Automatic Control & 2 & Automatica & 0.04478\\
2.372 &IEEE Control Systems Magazine & 3 & SIAM J. Control \& Optimization &0.01479\\
2.000 &IEEE Transactions Contr. Sys. Tech. & 4 & IEEE Transactions Contr. Sys. Tech. & 0.01196\\
1.900 &Int. J. Robust 
Nonlinear Control & 5 & Systems \& Control Letters &0.01087\\
1.805 &Journal Process Control & 6 & Int. Journal Control &0.00859\\
1.669 & Control Engineering Practice & 7 & Int. J. Robust Nonlinear Control&0.00854\\
1.667 &Systems \& Control Letters & 8 & Control Engineering Practice & 0.00696\\
1.379&SIAM J. Control \& Optimization & 9 & Journal Process Control&0.00622\\
1.250&European J. Control & 10 & IEEE Control Systems Magazine& 0.00554\\
\hline
\end{tabular}
\end{table}

\section*{Relations with Consensus Problems}

Consensus problems for multi-agent systems have a close relationship with
the PageRank problem and has motivated the distributed randomized approach 
introduced earlier. 
This section considers 
a stochastic version 
of the consensus problem, which has been studied
in, for example, \cite{BoyGhoPra:06,HatMes:05,TahJad:08,Wu:06};
see also \cite{TemIsh:07} for a discussion from the point of view of 
randomized algorithms. 
In \cite{TsuYam:08}, a dual version of PageRank is proposed and
its usefulness in controlling consensus-based agent systems is 
explored.

Consider a  network of $n$ agents corresponding to the vertices $\mathcal{V}=\{1,2,\ldots,n\}$ 
of a directed graph $\mathcal{G}=(\mathcal{V},\mathcal{E})$, where $\mathcal{E}$
is the set of links connecting the agents. The agent $i$ is said to be connected 
to the agent $j$ by a link $(i,j)\in\mathcal{E}$ if agent $i$ transmits its 
value to agent $j$. It is assumed that there exists a {\it globally reachable agent} 
for the  graph $\mathcal{G}$. This assumption implies that there exists an agent from which
every agent in the graph can be reached via a sequence of directed links, see, for example, 
\cite{BulCorMar:09,LinFraMag:05,RenBea:05}. Recall that the graph 
$\mathcal{G}$ has at least one globally reachable agent if and only if 1 is 
a simple eigenvalue of a row stochastic matrix representing the graph $I - L$, 
where $L$ is the Laplacian of $\mathcal{G}$ (for example, \cite{BulCorMar:09}).

The objective of consensus is that the values 
$x(k):= [x_1(k) \,\cdots\, x_n(k)]^T\in\R^n$ of all agents reach a common value 
by communicating to each other according to a prescribed communication pattern. 
Formally, consensus is said to be achieved in the sense of mean-square error (MSE) if, 
for any initial vector $x(0)\in\R^n$, it holds
\begin{equation}
\E\bigl[\abs{x_i(k)-x_j(k)}^2\bigr] \rightarrow 0,~~k\rightarrow\infty
\nonumber
\end{equation}
for all $i,j\in\mathcal{V}$.
The communication pattern (see ``Update Scheme for Consensus and Convergence Properties" 
for the precise definition) 
is determined at each time $k$ according to an i.i.d. random process
$\theta(k) \in\{1,\ldots,d\}$ with probability distribution given by 
$$
\Prob\{\theta(k)=i\} = \frac{1}{d},~~k\in\Z_+, 
$$
where $d$ is the number of patterns.


A classical approach used in the consensus literature is to update the value of each agent
by taking the average of the values received at that time.
The iteration can be written in the vector form as
\begin{equation}
x(k+1) = A_{\theta(k)} x(k),
\nonumber
\end{equation}
where the matrix $A_i$ is defined in
``Update Scheme for Consensus and Convergence Properties." 
In contrast to the PageRank case, only the agent values $x_i$ are updated,
and time averaging is not necessary for achieving probabilistic consensus.

\vspace{.5cm}

\centerline{
\fbox{\parbox{16cm}{
{\Large Update Scheme for Consensus and Convergence Properties}\\ \\
For each $i=1, 2, \ldots, d$, define a subset  $\mathcal{E}_i$ of the edge set $\mathcal{E}$ as follows:
\begin{enumerate}
\item[(i)]  For all $j \in \mathcal{V}$, $(j,j)\in\mathcal{E}_i$. \\
\vspace{-1cm}
\begin{equation}
\label{eqn:commpattern}
\end{equation}
\vspace{-1cm}
\item[(ii)]
$\bigcup_{i=1}^d \mathcal{E}_i=\mathcal{E}$.
\end{enumerate}
Let $\theta(k)$ be an independent and identically distributed (i.i.d.) random process
and its probability distribution is given by
\begin{equation}
  \Prob\{\theta(k)=i\} = \frac{1}{d},~~k\in\Z_+,
\nonumber
\end{equation}
where $d$ is the number of communication patterns.\\
Consider the update scheme
\begin{equation}
x(k+1) = A_{\theta(k)} x(k), 
\label{eqn:consensus:x}
\end{equation}
where the matrix $A_{i}$ is a row stochastic matrix constructed as follows
\[
 (A_{i})_{j\ell}
  := \begin{cases}
      \frac{1}{n_{ij}} & \text{if $(\ell,j)\in\mathcal{E}_{i}$},\\
      0                     & \text{otherwise}, 
    \end{cases}
\]
and $n_{ij}$ is the number of agents $\ell$ with $(\ell,j)\in\mathcal{E}_{i} \subseteq
\mathcal{E}$, that is, the number of agents sending information to agent $j$ under the 
communication pattern $\mathcal{E}_i$.\\
Assuming that a globally reachable agent exists, convergence of this scheme 
in the mean-square error (MSE) sense
\begin{equation}
\E\bigl[\abs{x_i(k)-x_j(k)}^2\bigr] \rightarrow 0,~~k\rightarrow\infty
\label{eqn:consensus}
\end{equation}
for all $i,j\in\mathcal{V}$; see for example,
\cite{IshTem:10,HatMes:05,TahJad:08, Wu:06}.
} 
}
}

\vspace{.5cm}

A simple example is now presented to illustrate the communication scheme. 

\begin{example}\label{ex:consensus}\rm
Consider the network of six agents illustrated in Fig.~\ref{fig:graph} (b) from Example~\ref{ex:1}. 
First, as a simple case, we look at a static communication scheme
where all agents communicate over the original edge set $\mathcal{E}$ at all times.
In this case, there is only one pattern $\mathcal{E}_1 = \mathcal{E}$ and hence $d=1$. 
To be consistent with the notation used for PageRank so far, 
the underlying matrix is simply denoted by $A$. This matrix is constructed 
using incoming (rather than outgoing) links to make it row stochastic as
\[
 A = \begin{bmatrix}
       \frac{1}{2} & \frac{1}{2} & 0           & 0           & 0           & 0           \\
       \frac{1}{3} & \frac{1}{3} & \frac{1}{3} & 0           & 0           & 0           \\
       0           & \frac{1}{3} & \frac{1}{3} & \frac{1}{3} & 0           & 0           \\
       \frac{1}{4} & 0           & \frac{1}{4} & \frac{1}{4} & 0           & \frac{1}{4} \\
       0           & 0           & 0           & \frac{1}{3} & \frac{1}{3} & \frac{1}{3} \\
       0           & 0           & \frac{1}{4} & \frac{1}{4} & \frac{1}{4} & \frac{1}{4} 
     \end{bmatrix}.
\]
This matrix is similar to the hyperlink matrix given in Example~\ref{ex:1} in the sense
that the nonzero off-diagonal entries coincide. However, this matrix is row stochastic
while the link matrix for PageRank is column stochastic.
Moreover notice that all diagonal entries of this matrix are positive,
resulting in the presence of self-loops in the graph.
(We recall that for PageRank no self-loops are considered 
because these loops may increase spamming; 
see further details in \cite{LanMey:06,AndBorHop:07}).


Next, as in the distributed case, we introduce six communication patterns 
arising from the protocol in the distributed PageRank algorithm.
The edge subset $\mathcal{E}_i$ contains all $(i,j)$ and $(j,i)$
links present in the original edge set $\mathcal{E}$ and all self-loops $(j,j)$ for $i,j=1,\dots,6$.
Then, the first three matrices $A_i$, $i=1,2,3$, are
\begin{align*}
 A_1 
  &= \begin{bmatrix}
       \frac{1}{2} & \frac{1}{2} & 0           & 0           & 0           & 0           \\
       \frac{1}{2} & \frac{1}{2} & 0           & 0           & 0           & 0           \\
       0           & 0           & 1           & 0           & 0           & 0           \\
       \frac{1}{2} & 0           & 0           & \frac{1}{2} & 0           & 0           \\
       0           & 0           & 0           & 0           & 1           & 0           \\
       0           & 0           & 0           & 0           & 0           & 1
     \end{bmatrix},~~
 A_2
   = \begin{bmatrix}
       \frac{1}{2} & \frac{1}{2} & 0           & 0           & 0           & 0           \\
       \frac{1}{3} & \frac{1}{3} & \frac{1}{3} & 0           & 0           & 0           \\
       0           & \frac{1}{2} & \frac{1}{2} & 0           & 0           & 0           \\
       0           & 0           & 0           & 1 & 0 & 0\\
       0           & 0           & 0           & 0 & 1 & 0\\
       0           & 0           & 0           & 0 & 0 & 1
     \end{bmatrix},~~
 A_3
   = \begin{bmatrix}
       1 & 0           & 0           & 0           & 0           & 0           \\
       0 & \frac{1}{2} & \frac{1}{2} & 0           & 0           & 0           \\
       0 & \frac{1}{3} & \frac{1}{3} & \frac{1}{3} & 0           & 0           \\
       0 & 0           & \frac{1}{2} & \frac{1}{2} & 0           & 0           \\
       0 & 0           & 0           & 0           & 1           & 0           \\
       0 & 0           & \frac{1}{2} & 0           & 0           & \frac{1}{2} 
     \end{bmatrix}.
\end{align*}
The rest of the matrices can be similarly obtained.

\End
\end{example}

\medskip



Table~\ref{consensusandpagerank}
summarizes some of the key differences and similarities between the consensus
problem addressed in this section and the distributed approach studied 
in ``Distributed Randomized Algorithms for PageRank Computation''
for the PageRank computation.

\begin{table}[htb]
\caption{Comparison between Consensus and PageRank}
\label{consensusandpagerank}
\begin{tabular}{|c|c|c|} \hline
 & Consensus & PageRank \\ \hline
objective & all agent values $x_i(k)$ become equal & page values $x_i(k)$ converge to a constant $x_i^*$\\
graph structure & a globally reachable agent exists & the web is not strongly connected \\
self-loops & presence of self-loops for agents & no self-loops are considered in the web\\
stochastic properties &row stochastic matrix $A$ & column stochastic matrices $A, M$\\
convergence & in mean-square error sense and with probability $1$ & in mean-square error sense
and with probability $1$
\\
initial conditions & convergence for any initial condition  $x(0) \in \R^n$ & convergence for stochastic vector $x(0) \in \R^n$\\
averaging & time averaging not necessary & time averaging $y(k)= \frac{1}{k+1}\sum_{\ell=0}^{k} x(\ell)$ required \\
\hline
\end{tabular}
\end{table}


\section*{Aggregation-Based PageRank Computation}

In this section, we turn our attention to a distributed PageRank computation
with a focus on reduced cost in computation and communication.
The particular approach developed in this section 
is based on the web aggregation technique proposed in 
\cite{IshTemBai:12}, which
leads us to a smaller web to be used in the computation. 
A procedure is presented to compute approximated values of the
PageRank and moreover provide an analysis on error bounds.
The approach shares ties with the
aggregation methods based on singular perturbation 
developed in the control literature \cite{PhiKok:81}.

\begin{figure}
  \centering
  \resizebox{6cm}{!}{\includegraphics{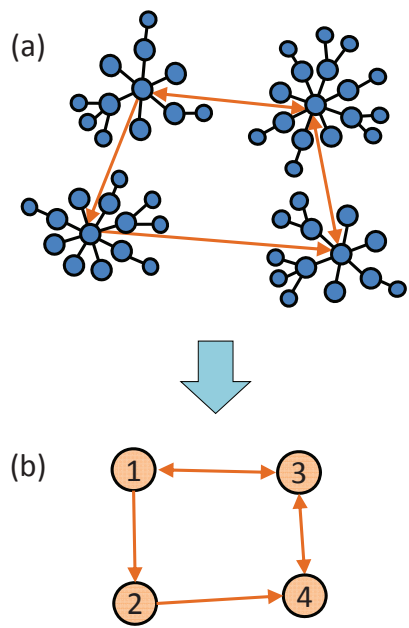}}
  \caption{A web graph with a sparse structure (a) and 
   its aggregated graph (b). It is known in the web that many links are internal, connecting pages
   within their own domain/directories. The aggregated web is obtained by grouping such pages.}
  \label{fig:aggregation1}  
\end{figure}

\subsection*{Aggregation of the Web}

The structure of the web is known to have a sparsity property because
many links are intra-host ones. This property means that pages are often linked 
within the same domains/directories (for example, organizations, universities/companies, 
departments, etc) \cite{LanMey:06,BroLem_infret:06}. 
A simple sparse network is illustrated in Fig.~\ref{fig:aggregation1} (a):
There are four domains with many pages, but the inter-domain links are only 
among the top pages.
In such a case, it is natural to group the pages and obtain an aggregated
graph with only four nodes as shown in Fig.~\ref{fig:aggregation1} (b).

By following this line of ideas, a PageRank computation approach
based on web aggregation is developed, roughly consisting of three steps: 
\begin{enumerate}
\item[1)] \textit{Grouping step}: Find groups in the web.
\item[2)] \textit{Global step}: Compute the total PageRank for each group.
\item[3)] \textit{Local step}: Distribute the total value of the group among members.
\end{enumerate}
The grouping of pages can mainly be done at the server level for the pages that
the servers host as we describe below.
The global step is at a higher level, requiring data exchange via communication among groups. 
In the local step, most of the computation should be carried out locally
within each group. 

For the aggregation-based method, pages are grouped 
with the purpose of computing the PageRank efficiently and accurately. 
Moreover, in view of the sparsity property of the web and the distributed algorithms 
discussed earlier, 
it is reasonable to group pages under the same servers or domains. This approach
has the advantage that grouping can be done in a decentralized manner.
More generally, the problem of grouping nodes in a network can be casted
as that of community detection, which can be performed based on different
measures such as modularity \cite{Newman:06,ExpEvaBlo:11} 
and the maximum flow \cite{FlaLawGil:02}. While such methods may be useful
for our purpose, they are known to be computationally expensive.

\begin{figure}
  \centering
  \resizebox{8cm}{!}{\includegraphics{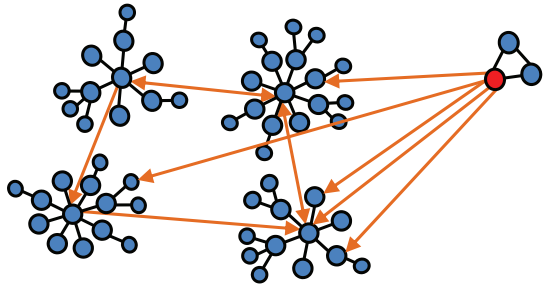}}
  \caption{A web page with many external links. In the real web, some pages have many
    outgoing links to pages outside of their own domains/directories. 
    In this respect, the sparsity property of the web is limited.}
  \label{fig:aggregation2}  
\end{figure}

From the viewpoint of a web page, the sparsity in the web structure can be
expressed by the limited number of links towards pages outside its own group.
That is, for page $i$, let its \textit{node parameter} $\delta_i\in[0,1]$ be given by
\begin{equation}
   \delta_i := \frac{\text{\# external outgoing links}}{\text{\# total outgoing links}}.
 \label{eqn:node_par}
\end{equation}
It is easy to see that smaller $\delta_i$ implies sparser networks. 
In the extreme case where the node parameters for \textit{all} pages are small as
\begin{equation}
   \delta_i \leq \delta~~\text{for each page $i$},
 \label{eqn:sparse_cond}
\end{equation}
where $\delta\in(0,1)$ represents the bound, 
then one may apply the aggregation techniques based on singular perturbation
for large-scale networked systems such as consensus and Markov chains
\cite{PhiKok:81,AldKha:91,BiyArc:08,ChoKok:85}.
However, in the real web, it is clear that pages with many external links
\textit{do} exist. Especially, if such pages belong to small groups, 
these results are not directly applicable (see Fig.~\ref{fig:aggregation2}).

In this aggregation approach, the main idea is to consider pages with many external links as
\textit{single groups} consisting of only one member. For such pages, the node parameters
always become 1.  Hence, such pages are excluded from the condition 
\eqref{eqn:sparse_cond} and instead we use the condition
\begin{equation}
   \delta_i \leq \delta~~\text{for each \textit{non-single} page $i$}.
 \label{eqn:sparse_cond2}
\end{equation}
The size of the parameter $\delta$ determines the number of groups
as well as the accuracy in the computation of PageRank. This point will
be demonstrated through the analysis of the approach and a numerical example. 

Denote by $r$ the number of groups and by $r_1$ the number of single groups. 
Also, for group $i$, let $\widetilde{n}_i$ be the number of member pages. 
For simplicity of notation, the page indices are reordered as follows:
In the PageRank vector $x^*$, the first $\widetilde{n}_1$ 
elements are for the pages in group $1$, and the following $\widetilde{n}_2$ entries are 
for those in group $2$, and so on.

\subsection*{Approximated PageRank via Aggregation}

For group $i$, its \textit{group value} denoted by $\widetilde{x}_{1i}^*$ 
is given by the total value of the PageRanks $x_j^*$ of its members. 
Hence, introduce a coordinate transformation as
\begin{equation*}
\widetilde{x}^*:=Vx^*
~~\Leftrightarrow~~
     \begin{bmatrix}
        \widetilde{x}_1^*\\
        \widetilde{x}_2^*
     \end{bmatrix}
   := \begin{bmatrix}
         V_1\\ V_2
      \end{bmatrix} x^*,
  \label{eqn:xtilde}
\end{equation*}
where
\begin{itemize}
 \item $\widetilde{x}_1^*\in\R^{r}$: the $i$th entry is the group value of group $i$;
 \item $\widetilde{x}_2^*\in\R^{n-r}$: each entry represents the difference between 
                                   a page value and the average value of the group members.
\end{itemize}
The first vector $\widetilde{x}_1^*$ is called the \textit{aggregated PageRank}.
By definition, it follows that $\widetilde{x}_1^*\geq 0$ and
$\one_r^T\widetilde{x}_1^*=1$.

The transformation matrix $V = \big[V_1^T~V_2^T\big]^T\in\R^{n\times n}$ is
\begin{align}
\begin{split}
 V_1 &:= \text{bdiag}
            (\one_{\widetilde{n}_i}^T)
            \in\R^{r\times n},\\
 V_2 &:= \text{bdiag}
         \Big(
          [I_{\widetilde{n}_i-1}~0]
            - \frac{1}{\widetilde{n}_i} 
               \one_{\widetilde{n}_i-1}\one_{\widetilde{n}_i}^T
         \Big)
         \in\R^{(n-r)\times n},
\end{split}
\label{eqn:V}
\end{align}
where $\text{bdiag}(X_i)$ denotes a block-diagonal matrix
whose $i$th diagonal block is $X_i$.
The matrices $V_1$ and $V_2$ are block diagonal,
containing $r$ and $r-r_1$ blocks, respectively. 
Note that in $V_2$, if the $i$th group is a single one
(that is, $\widetilde{n}_i=1$), then the $i$th block has the size
$0\times 1$, meaning that the corresponding column is zero. 
Due to the simple structure, the inverse of this matrix $V$ 
can be obtained explicitly \cite{IshTemBai:12}.


Now, the definition of PageRank in \eqref{eqn:prvec} can 
be rewritten for $\widetilde{x}^*$ in the new coordinate as
\begin{align}
 \begin{bmatrix}
   \widetilde{x}_1^*\\
   \widetilde{x}_2^*
 \end{bmatrix}
  &= (1-m)\begin{bmatrix}
             \widetilde{A}_{11} & \widetilde{A}_{12}\\
             \widetilde{A}_{21} & \widetilde{A}_{22}
          \end{bmatrix}
           \begin{bmatrix}
             \widetilde{x}_1^*\\
             \widetilde{x}_2^*
           \end{bmatrix}
      + \frac{m}{n}
           \begin{bmatrix}
             u\\
             0
           \end{bmatrix}
  \label{eqn:Atil0}
\end{align}
with $\widetilde{A}_{11}\in\R^{r\times r}$ and
$u:=V_1\one_n=[\widetilde{n}_1\;\cdots\widetilde{n}_r]^T$.
As explained in ``Approximated PageRank Computation,"
the expression \eqref{eqn:Atil0} has two 
important properties: (i)~The (1,1)-block matrix $\widetilde{A}_{11}$ is 
a stochastic matrix, and (ii)~the entries of $\widetilde{A}_{12}$ are
``small'' in magnitude due to the sparse structure of the web. 

These properties form the basis for introducing an approximate version of
PageRank by triangonalizing the matrix $A$ in \eqref{eqn:Atil0} as follows: 
Let $\widetilde{x}_1'\in\R^{r}$ and $\widetilde{x}_2'\in\R^{n-r}$ be
the vectors satisfying 
\begin{align}
 \begin{bmatrix}
   \widetilde{x}_1'\\
   \widetilde{x}_2'
 \end{bmatrix}
  &= (1-m)\begin{bmatrix}
             \widetilde{A}_{11} & 0\\
             \widetilde{A}_{21} & \widetilde{A}_{22}'
          \end{bmatrix}
           \begin{bmatrix}
             \widetilde{x}_1'\\
             \widetilde{x}_2'
           \end{bmatrix}
      + \frac{m}{n}
           \begin{bmatrix}
             u\\
             0
           \end{bmatrix},
  \label{eqn:Atil0dash}
\end{align}
where $\widetilde{x}_1'$ is a probability vector.
The $(2,2)$-block matrix $\widetilde{A}_{22}'$
is taken as a block-diagonal matrix in accordance with the grouping;
for more details, see ``Approximated PageRank Computation."

Let $\widetilde{x}':=[\widetilde{x}_1'^T~\widetilde{x}_2'^T]^T$.
Then, the \textit{approximated} PageRank is obtained in the original coordinate as
\begin{equation}
   x':= V^{-1}\widetilde{x}'.
 \label{eqn:xdash0} 
\end{equation}

\subsection*{Computation of Approximated PageRank}

This section outlines an algorithm, consisting of three steps, for computing the approximated PageRank.

\begin{algorithm}
\label{alg:1}\rm 
Take the initial state $\widetilde{x}_1(0)\in\R^r$ as a stochastic vector, and 
then proceed according to the following three steps:

1.~Iteratively, compute the first state $\widetilde{x}_1(k)\in\R^r$ by
\begin{align}
 \widetilde{x}_1(k+1)
  &= (1-m)\widetilde{A}_{11} \widetilde{x}_1(k)
       + \frac{m}{n}u.
\label{eqn:updateVx_red1}
\end{align}

2.~After finding the first state $\widetilde{x}_1(k)$,
compute the second state $\widetilde{x}_2(k)\in\R^{n-r}$ by
\begin{align}
 \widetilde{x}_2(k)
  &= (1-m)\big[
            I - (1-m)\widetilde{A}'_{22}
          \big]^{-1}
           \widetilde{A}_{21} \widetilde{x}_1(k).
\label{eqn:updateVx_red2}
\end{align}

3.~Transform the state back in 
the original coordinate by
\begin{equation}
 x(k) = V^{-1} \begin{bmatrix}
                   \widetilde{x}_1(k)\\
                   \widetilde{x}_2(k)
               \end{bmatrix}.
 \label{eqn:updateVx_red3}
\end{equation}
\end{algorithm}

\medskip
The first and second steps in the algorithm are based on the
definition of $\widetilde{x}'$ in \eqref{eqn:Atil0dash}.
It requires the recursive computation of only the first state 
$\widetilde{x}_1(k)$, whose dimension 
equals the number $r$ of groups. At this stage, 
information is exchanged only among the groups.
By \eqref{eqn:Atil0dash}, the second state $\widetilde{x}_2'$ can
be computed recursively through
\[
 \widetilde{x}_2(k+1)
  = (1-m)\widetilde{A}_{21} \widetilde{x}_1(k)
     + (1-m) \widetilde{A}'_{22} \widetilde{x}_2(k).
\]
Here, we employ the steady state form, that is, $\widetilde{x}_2(k+1)=\widetilde{x}_2(k)$, 
to obtain the equation in \eqref{eqn:updateVx_red2}.
Note that the matrix $I - (1-m)\widetilde{A}_{22}'$ is nonsingular
because $(1-m)\widetilde{A}_{22}'$ is a stable matrix (as seen in 
``Approximated PageRank Computation").
This approach is motivated by the time-scale separation in methods based on
singular perturbation.
The update scheme in the algorithm is guaranteed to converge to 
the approximated PageRank vector $x'$, which follows from \eqref{eqn:conv}
\[
  x(k)\rightarrow x'~~\text{as}~k\rightarrow\infty.
\]

It is observed that this algorithm is suitable for distributed computation. 
The first step \eqref{eqn:updateVx_red1} is the global step
where the groups communicate to each other and
compute the total of the members' values, represented by 
the $r$-dimensional state $\widetilde{x}_1(k)$.
It is important to note that this step can be implemented via the 
distributed randomized approach discussed in 
``Distributed Randomized Algorithms for PageRank Computation.''
The rest of the algorithm can be carried out mostly  
via local interactions within each group.
This local computation can be confirmed from the block-diagonal structures in
the matrices $I - (1-m)\widetilde{A}_{22}'$ and $V^{-1}$.
The only part that needs communication over inter-group links
is in the second step \eqref{eqn:updateVx_red2}, when 
computing the vector $\widetilde{A}_{21} \widetilde{x}_1(k)$.

\begin{table}[t]
\begin{center}
\caption{Comparison of operation costs with communication among groups}
\label{table:operation}
\begin{tabular}{lll}
  \hline
   Algorithm & Equation & Bound on numbers of operations\\
  \hline
  Original & \eqref{eqn:xM0} 
    & $O((2 f_{0}(A)+n)\overline{k}_1)$\\
  Aggregation based & \eqref{eqn:updateVx_red1}
    & $O((2f_{0}(\widetilde{A}_{11})+r)\overline{k}_2$\\
    & &  \hspace*{1.1cm}~~$\mbox{} + f_{0}(A_{\text{ext}}) + n + r)$\\
                    & \eqref{eqn:updateVx_red2}
    & $O(2 f_{0}(A)+2n+r)$\\
  \hline
\end{tabular}
\end{center}
\hspace*{6cm} 
$f_0(\cdot)$: The number of nonzero entries of a matrix\\[.5mm]
\hspace*{6cm} 
$\overline{k}_1,\overline{k}_2$: The numbers of steps in the recursions
\end{table}

We now discuss the computational advantages of 
the aggregation-based approach of Algorithm~\ref{alg:1} over
the original scheme \eqref{eqn:xM0}. 
The number of operations for both schemes are displayed in 
Table~\ref{table:operation}.
Here, $f_{0}(A)$ denotes the number of nonzero entries
in the link matrix $A$. For the product of a sparse matrix $A$ and
a vector, operations of order $2f_{0}(A)$ are necessary.
The numbers of steps for the schemes to converge are denoted by
$\overline{k}_1$ and $\overline{k}_2$.
For termination criteria, see, for example, \cite{KamHavGol:04} for the centralized case 
and \cite{IshTem:10} for the distributed case. 

For the aggregation-based scheme, the table shows the operational cost
involving communication among groups. Other computation is decentrally 
carried out and is of polynomial orders of $n_i$ for group $i$.
The first step \eqref{eqn:updateVx_red1}
requires the computation of $\widetilde{A}_{11}$ and the iteration. 
For the second step \eqref{eqn:updateVx_red2}, 
we counted the multiplication of $\widetilde{A}_{21}\widetilde{x}_{1}(k)$.
As discussed earlier, 
the matrix $\widetilde{A}_{22}'$ is block diagonal, whose blocks
are of the size $(n_i-1)\times(n_i-1)$. The inverse of each block
can be obtained at the corresponding group. 
The same holds for the third step \eqref{eqn:updateVx_red3},
where the inverse $V^{-1}$ has a block-diagonal like structure
that can be easily obtained analytically.



\fbox{
\begin{minipage}{16cm}
{\Large Approximated PageRank Computation}\\ \\
%
The key observation in the proposed aggregation is that
in the original definition of PageRank vector $x^*$, 
the hyperlink matrix $A$ can be decomposed into three parts as
\begin{center}
  \resizebox{9cm}{!}{\includegraphics{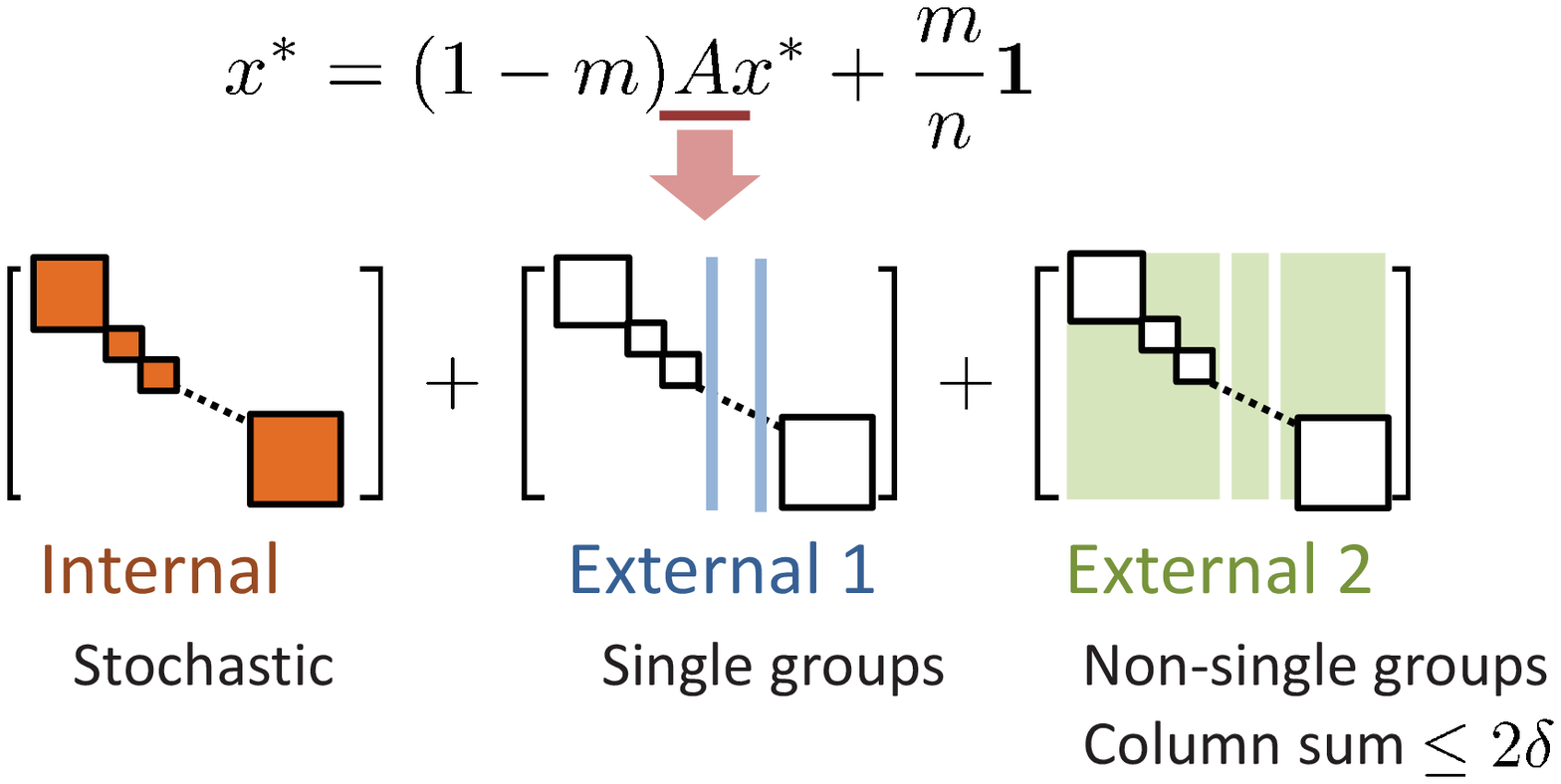}}
  \vspace*{-6mm}
\end{center}
where the elements in the colored area are nonzero.
The three matrices can be described as follows.
\begin{enumerate}
\item[(i)] The \textit{internal} link matrix (orange):
This matrix represents only the links within groups. It is block diagonal 
where the $i$th block is $\widetilde{n}_i\times \widetilde{n}_i$,
corresponding to group $i$.
Its nondiagonal entries are the same as those of $A$, but its diagonal entries 
are chosen so that this matrix becomes stochastic. 
\item[(ii)] The \textit{external} link matrix 1 (light blue):
This matrix contains the columns of $A$ corresponding to the single groups.
Its diagonal entries are changed to make the column sums zero.
\item[(iii)] The \textit{external} link matrix 2 (light green): 
The remaining columns for the non-single groups are put together here,
and thus this matrix has a sparse structure. 
In fact, based on the bound $\delta$
on the node parameters in \eqref{eqn:sparse_cond2}, the column sums are
bounded by $2\delta$. 
(The diagonal elements are chosen as in the previous matrix.)
\end{enumerate}
After the change in the coordinate, the PageRank vector $\widetilde{x}^*$
can be expressed as 
\begin{center}
  \resizebox{9cm}{!}{\includegraphics{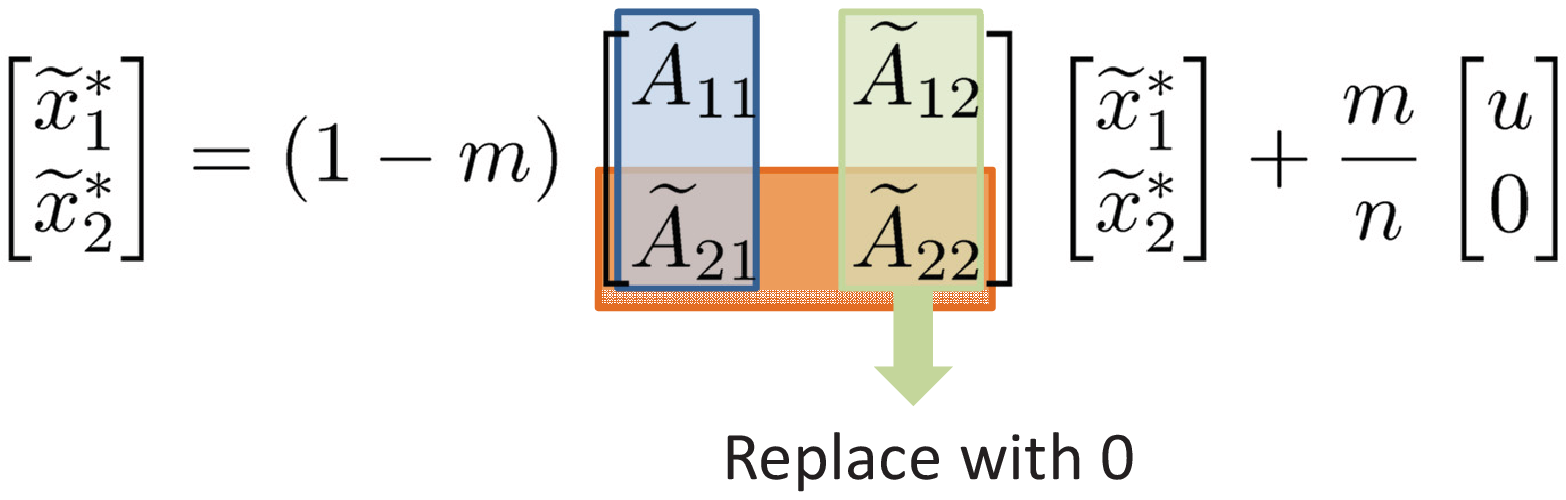}}
  \vspace*{-6mm}
\end{center}
where in colors, the contributions of the three parts of $A$ 
to the submatrices $\widetilde{A}_{ij}$ are indicated. We remark that the internal
links only appear in $\widetilde{A}_{21}$ and $\widetilde{A}_{22}$. 
The external link matrices~1 and 2 also affect different parts of the matrix. 
An important aspect is that the external link matrix~2 for non-single groups
(light green), which has only small entries outside the diagonal blocks,
contributes to the submatrices $\widetilde{A}_{12}$ and $\widetilde{A}_{22}$.\\
\end{minipage}
}

\fbox{
\begin{minipage}{16cm}
\vspace*{0.5cm}
Thus, the approach in approximating the PageRank requires replacing its contribution 
with zero. We then arrive at the definition of $\widetilde{x}'$
\begin{center}
  \resizebox{9cm}{!}{\includegraphics{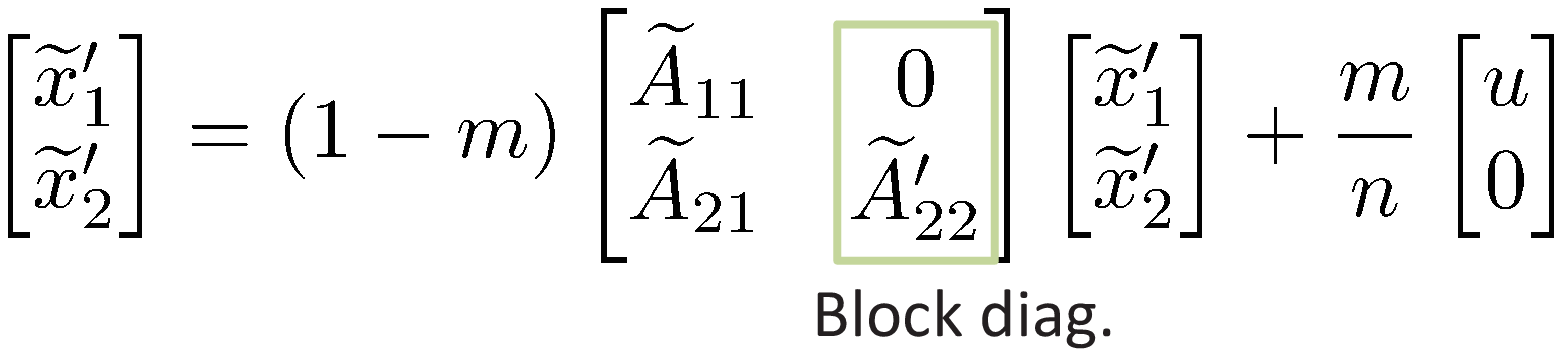}}
  \vspace*{-6mm}
\end{center}
where the (1,2)-block is zero, and 
the (2,2)-block submatrix $\widetilde{A}_{22}'$ is block diagonal. From
this vector, the approximated PageRank $x'$ in \eqref{eqn:xdash0} is obtained. 

The following properties are important to justify the aggregation-based approach:
 \begin{enumerate}
 \item[1.] The matrix $\widetilde{A}_{11}$ is stochastic. 
 \item[2.] The matrix $\widetilde{A}'_{22}$ 
       has spectral radius smaller than or equal to 1.
 \item[3.] The approximated PageRank vector $x'$ exists and is unique.
\end{enumerate} 
\vspace*{0.5cm}
\end{minipage}
}


\subsection*{Analysis of Approximation Error}

The error in the approximated PageRank just introduced is examined here.
It turns out that the error can be related to
the level of sparsity when aggregating the web, 
represented by the node parameter $\delta$ in \eqref{eqn:sparse_cond}.

For a given $\epsilon \in (0,1)$, 
to achieve an error between the approximated PageRank $x'$ in \eqref{eqn:xdash0} 
and the PageRank vector $x^*$ expressed by $\epsilon$ as
\begin{equation*}
   \norm{x^* - x'}_1 \leq \epsilon,
\end{equation*}
it is sufficient that the web is aggregated so that 
the node parameter $\delta$ is small enough that
\begin{equation*}
   \delta 
       \leq \frac{m\epsilon}{4(1-m)(1+\epsilon)}.
   \label{eqn:delta2}
\end{equation*}

This result \cite{IshTemBai:12} shows that if the web can be aggregated so that 
the ratio of external outgoing links (represented by $\delta$) 
is limited for non-single pages,
then a good approximate of the PageRank can be computed
through a lower-order algorithm. It will be shown through
a numerical example later that a tradeoff exists between
the size of $\delta$ and the number $r$ of groups, which
directly influences the computation load for the algorithm.

\begin{example}
\rm
\label{ex:orig}

\begin{figure}
  \centering
  \resizebox{9cm}{!}{\includegraphics{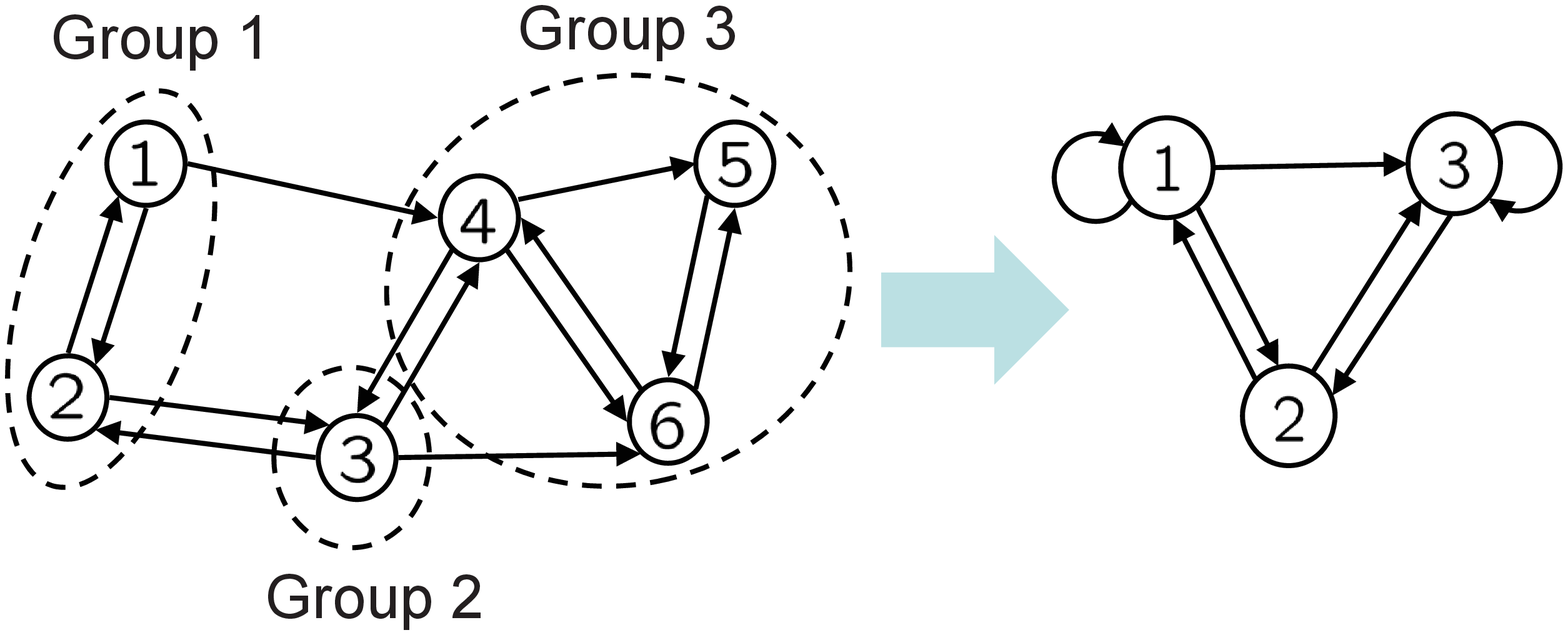}}
  \caption{The six-page web partitioned into three groups (a) and its aggregated graph (b).
       The web pages in the original graph are grouped in such a way that the numbers of 
       external links towards other groups are small. Group~2 consists of only one page, making
       it a single group.}           
  \label{fig:agg}  
\end{figure}

We continue with the six-page web example to examine the approximated
PageRank based on the aggregated web. 

First, the pages are partitioned into three groups as shown in Fig.~\ref{fig:agg} (a).
In the aggregated graph in Fig.~\ref{fig:agg} (b), 
nodes 1 and 3 have self-loops.

\begin{table}[t]
\begin{center}
\caption{Ratios of external links for the six pages in Fig.~\ref{fig:agg}}
\label{table:nodepara}
\begin{tabular}{ccccccc}
  \hline\\[-4mm]
    Page Index & 1 & 2 & 3 & 4 & 5 & 6\\[1mm]
  \hline\\[-4mm]  
   {\large $\frac{\text{\# external outlinks}}{\text{\# outlinks}}$}
      & $\frac{1}{2}$ & $\frac{1}{2}$ & 1 & $\frac{1}{3}$ & 0 & 0\\[1mm]
  \hline
\end{tabular}
\end{center}
\end{table}

As seen in Table~\ref{table:nodepara},
after this aggregation, the ratio of external links is limited for each page 
except for page~3 which forms a single group. 
Hence, the node parameter in \eqref{eqn:sparse_cond} is taken to be $\delta=0.5$.
The matrix $V$ in \eqref{eqn:V} for the coordinate change is defined as

\vspace*{-2mm}
{\small
\[
 V = \left[
       \begin{array}{c}
         V_1\\       
         \hline
         V_2
       \end{array}
     \right]
   = \left[
     \begin{array}{ccccccc}
       \cdashline{1-2}
       \multicolumn{1}{:c}{1} & \multicolumn{1}{c:}{1} & 0 & 0 & 0 & 0\\
       \cdashline{1-3}
       0 & 0 & \multicolumn{1}{:c:}{1} & 0 & 0 & 0\\
       \cdashline{3-6}
       0 & 0 & 0 & \multicolumn{1}{:c}{1} & 1 & \multicolumn{1}{c:}{1}\\
       \cdashline{4-6}\\[-3mm]
       \hline\\[-3mm]
       \cdashline{1-2}
       \multicolumn{1}{:c}{1/2} & \multicolumn{1}{c:}{-1/2} & 0   & 0 & 0 & 0\\
       \cdashline{1-2}\cdashline{4-6}
       0 & 0 & 0 &  \multicolumn{1}{:c}{2/3} & -1/3 & \multicolumn{1}{c:}{-1/3}\\
       0 & 0 & 0 & \multicolumn{1}{:c}{-1/3} &  2/3 & \multicolumn{1}{c:}{-1/3}\\
       \cdashline{4-6}
    \end{array}\right],
\]}

\vspace*{-2mm}
\noindent
where the dashed-line boxes indicate the diagonal blocks.
Since group~2 is single, the third column in $V_2$ is zero. 

After the change in its coordinate,
the PageRank vector $\widetilde{x}^*=V x^*$ becomes
\begin{align*}
 \widetilde{x}^* 
  &= \left[
       \begin{array}{c|c}
         (\widetilde{x}_1^*)^T & (\widetilde{x}_2^*)^T  
       \end{array}
    \right]^T
  = \left[
       \begin{array}{ccc|ccc}
         0.147 & 0.122 & 0.731 & -0.0121 &  -0.0294 & -0.0294
       \end{array}
    \right]^T.    
\end{align*}
Note that the entries of the first part $\widetilde{x}_1^*$ sum up to 1.

According to the discussion in ``Approximated PageRank Computation," 
we can decompose the link matrix $A$ into three parts as follows:
\vspace*{-2mm}
{\small
\begin{align*}
 A &= \text{(Internal)} + \text{(External 1)} + \text{(External 2)}\\
  &= \left[
     \begin{array}{cccccc}
      \cdashline{1-2}
      \multicolumn{1}{:c}{1/2} & \multicolumn{1}{c:}{1/2}  &  0  & 0 &  0  &  0\\
      \multicolumn{1}{:c}{1/2}  & \multicolumn{1}{c:}{1/2} &  0  & 0 &  0  &  0\\
      \cdashline{1-3}
         0 &  0   &  \multicolumn{1}{:c:}{1}  & 0 &  0  &  0\\
      \cdashline{3-6}
         0 &  0   &  0  & \multicolumn{1}{:c}{1/3} & 0 & \multicolumn{1}{c:}{1/2}\\
         0 &  0   &  0  & \multicolumn{1}{:c}{1/3} & 0 & \multicolumn{1}{c:}{1/2}\\
         0 &  0   &  0  & \multicolumn{1}{:c}{1/3} &  1 & \multicolumn{1}{c:}{0} \\
      \cdashline{4-6}
     \end{array}
    \right]
   + 
    \left[
     \begin{array}{cccccc}
      \cdashline{1-2}
      \multicolumn{1}{:c}{0} &  \multicolumn{1}{c:}{0}  &  0   &  0   &  0  &  0\\
      \multicolumn{1}{:c}{0}   & \multicolumn{1}{c:}{0} &  1/3 &  0   &  0  &  0\\
      \cdashline{1-3}
       0   & 0 & \multicolumn{1}{:c:}{-1}   &  0 &  0  &  0\\   
      \cdashline{3-6}
         0 &  0   &  1/3 & \multicolumn{1}{:c}{0} &  0  & \multicolumn{1}{c:}{0}\\   
         0 &  0   &  0   & \multicolumn{1}{:c}{0}    &  0  & \multicolumn{1}{c:}{0}\\
         0 &  0   &  1/3 & \multicolumn{1}{:c}{0}    &  0  & \multicolumn{1}{c:}{0}\\
      \cdashline{4-6}
     \end{array}
    \right]
    + 
    \left[
     \begin{array}{cccccc}
      \cdashline{1-2}
      \multicolumn{1}{:c}{-1/2} &  \multicolumn{1}{c:}{0}  &  0   &  0   &  0  &  0\\
      \multicolumn{1}{:c}{0}   & \multicolumn{1}{c:}{-1/2} &  0   &  0   &  0  &  0\\
      \cdashline{1-3}
       0   & 1/2 & \multicolumn{1}{:c:}{0}   &  1/3 &  0  &  0\\   
      \cdashline{3-6}
       1/2 &  0   &  0 & \multicolumn{1}{:c}{-1/3} &  0  & \multicolumn{1}{c:}{0}\\   
         0 &  0   &  0   & \multicolumn{1}{:c}{0}    &  0  & \multicolumn{1}{c:}{0}\\
         0 &  0   &  0 & \multicolumn{1}{:c}{0}    &  0  & \multicolumn{1}{c:}{0}\\
      \cdashline{4-6}
     \end{array}
    \right].
\end{align*}}

\vspace*{-2mm}
\noindent
The internal part contains the block-diagonal entries of $A$
and is a stochastic matrix. 

In the new coordinates, 
the link matrix $\widetilde{A}$ in \eqref{eqn:Atil0} is

\vspace*{-2mm}
{\small
\begin{align*}
  & \widetilde{A}
     = \left[
         \begin{array}{c|c}
          \widetilde{A}_{11} & \widetilde{A}_{12}\\
          \hline
          \widetilde{A}_{21} & \widetilde{A}_{22}
         \end{array}
       \right]
   = \left[
         \begin{array}{ccc|ccc}
           0.5  & 0.333 & 0     &  0   &  0     & 0\\
           0.25 & 0     & 0.111 & -0.5 &  0.333 & 0\\
           0.25 & 0.667 & 0.889 &  0.5 & -0.333 & 0\\
           \hline
           0   & -0.167 &      0 &  -0.5  &        0  &       0\\
           0.167 &  0.111 & -0.130 &   0.333 &   -0.389 &   -0.5\\   
          \hspace*{-1mm}-0.0833 \hspace*{-1mm}&  -0.222 \hspace*{-1mm}&  -0.0185 
                & -0.167\hspace*{-1mm} &   -0.0556\hspace*{-1mm} &   -0.5\hspace*{-1mm}
	     \end{array}
	  \right].
\end{align*}}

\vspace*{-2mm}
\noindent
For computation of the approximated PageRank, 
Algorithm~\ref{alg:1} uses the matrices $\widetilde{A}_{11}$ above and

\vspace*{-2mm}
{\small
\begin{align*}
  \begin{bmatrix}
    I - (1-m)\widetilde{A}'_{22}
  \end{bmatrix}^{-1}
    \widetilde{A}_{21}
     = \begin{bmatrix}
         0 & -0.167 & 0\\
         0.174 & 0.161 & -0.113\\
         -0.0758 & -0.172 & -0.00177
       \end{bmatrix},
\end{align*}}

\vspace*{-2mm}
\noindent
where $\widetilde{A}'_{22}$ is given by

\vspace*{-2mm}
{\small
\begin{align*}
  \widetilde{A}'_{22}   
    = \left[\begin{array}{ccc}
        \cdashline{1-1}
        \multicolumn{1}{:c:}{0} &   0     &   0\\
        \cdashline{1-3}
        0 &  \multicolumn{1}{:c}{-0.167} &  \multicolumn{1}{c:}{-0.5}\\
        0 &  \multicolumn{1}{:c}{-0.167} &  \multicolumn{1}{c:}{-0.5}\\
        \cdashline{2-3}
      \end{array}\right].
\end{align*}}

\vspace*{-2mm}
\noindent
It should be noted that $\widetilde{A}_{11}$ is stochastic. Further, 
$\widetilde{A}_{22}'$ is block diagonal, which is a property
not shared with $\widetilde{A}_{22}$, and is also Schur stable.

Now, the approximated PageRank vector is
\begin{align*}
  x' &= V^{-1} \widetilde{x}'
      = \begin{bmatrix}
         0.0566 & 0.0920 & 0.125 & 0.212 & 0.213 & 0.302
       \end{bmatrix}^T.
\end{align*}
The difference between this vector and the original PageRank $x^*$ 
in \eqref{eqn:ex:x_ast} is certainly small 
\[
 \norm{x'-x^*}_1 
    = 0.0188.
\]
\End
\end{example}

\section*{Experimental Results}

In this section, numerical simulations are presented based on 
a web obtained from real data.

\begin{figure}[t]
  \centering
  \fig{13cm}{!}{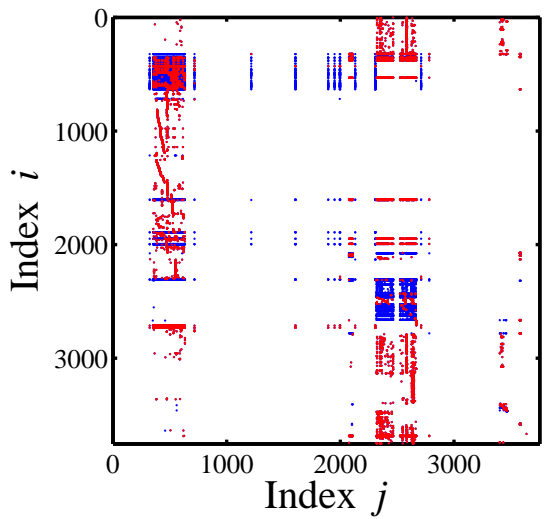}
  \caption{The graph structure of the web data used in experiments. 
   The points indicate links from page $j$ to page $i$ and the red points 
   are those linking to dangling nodes.
   Two clusters of pages with dense link structures can be found around 
   indices 500 and 2,500.}
  \label{fig:G_orig}
\end{figure}

\begin{figure}[t]
  \centering
  \fig{9cm}{!}{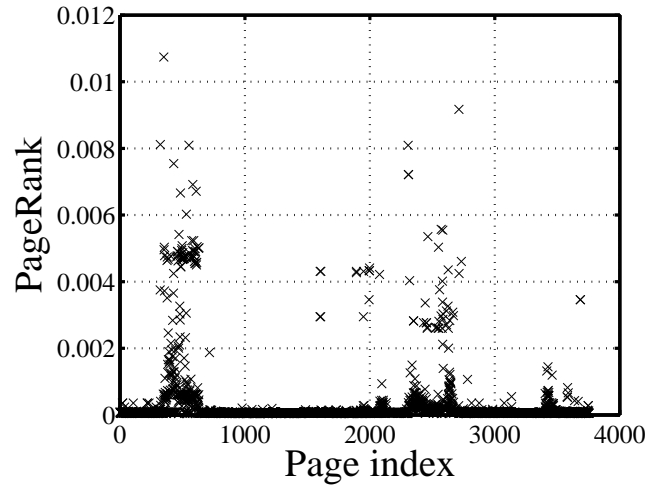}
  \caption{PageRank values of all pages in the example web. The values for those pages in the
   two clusters (around indices 500 and 2,500) are especially high.}
  \label{fig:PageRank_values}
\end{figure}

\subsection*{The Web Data and its PageRank}

First, we describe the web data that has been employed in this simulation. 
The data was obtained from the database \cite{webdata} collected by crawling 
web pages of various universities.
This database has previously been used as a benchmark for testing PageRank algorithms \cite{FABG:13}.
Among them, we have chosen the data from Lincoln University in New Zealand from the year 2006.
This web has 3,756 nodes
with 31,718 links and there are in total 684 domains. 
The largest is the main domain of the university (www.lincoln.ac.nz), consisting of
2,467 pages. Other larger domains in this dataset contained
221, 101, 68, 24 pages, and so on. 
In the real web, 
a fairly large portion of the nodes are dangling nodes.
In this example, there are 3,255 dangling nodes, which is over 85 percent
of the total.
Also, two nodes had no incoming links; these were removed since 
such nodes play very minor roles in the PageRank values.
The pages were indexed according to the domain/directory names in an alphabetic order.
Fig.~\ref{fig:G_orig} displays the link pattern of the web with $n=3,754$ pages,
where the blue points represent the nonzero entries of the connectivity matrix;
the red points correspond to outgoing links from dangling nodes.

To proceed with the PageRank computation, the web needs to be modified
so that the resulting link matrix $A$ becomes stochastic. 
This modification was done by adding back links to dangling nodes, that is, links from each dangling 
node to the pages that have links to it. 
Hence, in the link pattern of Fig.~\ref{fig:G_orig},
for each red points in the $(i,j)$th entry, a new point in the $(j,i)$th entry was added.
The resulting web had 40,646 links.
For this web, the PageRank values were calculated by the power method. 
About 40 iterations were sufficient for its convergence. 
The results are shown in Fig.~\ref{fig:PageRank_values}. Comparing this with the
link structure in Fig.~\ref{fig:G_orig}, we notice that the pages with higher PageRank
values are included in the two clusters where many pages are linked to each other,
especially around page indices 500 and 2,500.
The top two pages in PageRank values turned out to be the ``search'' pages of the university
while the main home page of the university came in the third place.

\subsection*{Distributed Randomized Algorithm}

These values could also be computed via the distributed randomized algorithm.
Here, we use a modified version of the algorithm from 
``Distributed Randomized Algorithms for PageRank Computation''
based on the simultaneous updates \cite{IshTem:10}. 
In contrast with the original scheme,  
at each time step,
each page asynchronously decides to send its value to its neighbors 
in a probabilistic way under a fixed probability. 
Thus, even in the event that an agent receives data from
multiple agents at the same time, this algorithm can handle all data in
the update at that moment.
Another benefit is that the convergence is faster. 
Throughout this section, this update
probability is fixed to be 0.2, so on average each agent makes a transmission
once in every five time steps. 

\begin{figure}[t]
  \centering
  \fig{9cm}{!}{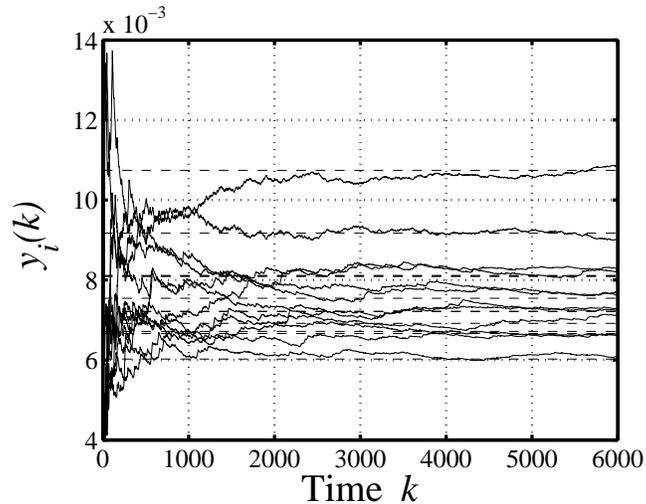}
  \caption{Time responses of the time averages $y_i(k)$ for some pages and
       their corresponding PageRank values (in dashed lines).}
  \label{fig:timeresp_y}
\end{figure}
\begin{figure}[t]
  \centering
  \fig{9cm}{!}{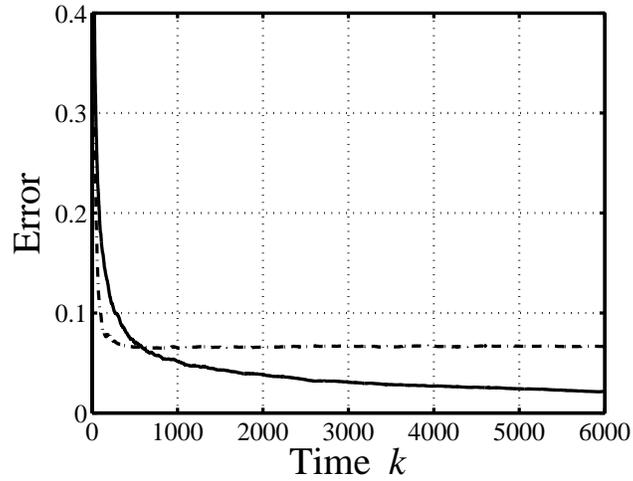}
  \caption{Time responses of the error from the PageRank $x^*$ in the distributed randomized algorithm 
             for the full-order case (solid) and the aggregation-based case (dash-dot).
             The response of the aggregation-based scheme is faster, but some error remains
             since the approximated PageRank $x'$ is computed.}
  \label{fig:timeresp_err1}
\end{figure}

In Fig.~\ref{fig:timeresp_y}, the time averages $y_i(k)$
are displayed for the pages taking larger values of PageRank. The true PageRank values
are indicated in dashed lines, and the convergence to these lines is observed.
Moreover, to see the overall convergence rate, we plotted in Fig.~\ref{fig:timeresp_err1}
the response of the error $\|y(k)-x^*\|_1$ from the true values in 1-norm (solid line).

\subsection*{Aggregation-Based Computation}

\begin{figure}[t]
  \centering
  \fig{9cm}{!}{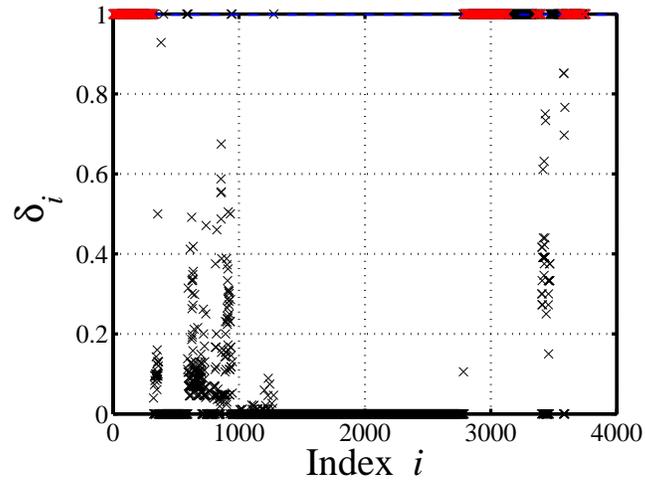}
  \caption{The node parameters $\delta_i$ with the original grouping based on domains.
           Pages taking large values are mostly single groups (marked in red), 
           but among the pages in non-single groups (marked in black), some have 
           a large portion of external links.}
  \label{fig:delta_i_1}
\end{figure}

We further continued with computation based on the technique of web aggregation
from ``Aggregation-Based PageRank Computation.''
The first step is to specify the groups of pages, from which we can estimate the 
sparsity structure in the web based on the node parameters $\delta_i$ in \eqref{eqn:node_par}; 
for page $i$, this parameter $\delta_i$ 
indicates the fraction of internal links within its own group over all of its outlinks. 
A simple way to find the initial grouping is to divide the pages based on their domain names.
Fig.~\ref{fig:delta_i_1} shows the node parameters for all pages based on this initial grouping. 
Each mark in red indicates a page which has no other page in its domain and 
hence is identified as a single group. Such pages necessarily have node parameters of 1, and
there were 577 of them. 
Note that this grouping resulted in a limited 
number of pages in non-single groups taking large values of $\delta_i$. 
However, some of them have $\delta_i=1$, meaning that the aggregation method is not
directly applicable at this point. 

\begin{figure}[t]
  \centering
  \fig{9cm}{!}{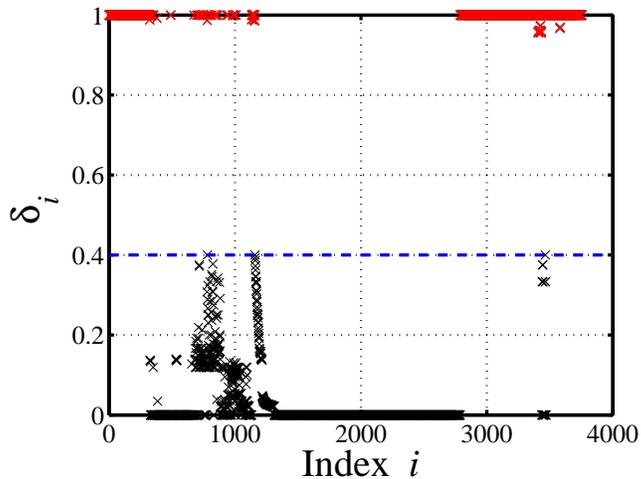}
  \caption{The node parameters $\delta_i$ after regrouping with $\delta=0.4$.
     Pages taking large values are all single groups (marked in red). Other pages
     are grouped so that their node parameters remain below the threshold $\delta$.}
  \label{fig:delta_i_04}
\end{figure}

Therefore, the next step is to remove pages taking larger values from their groups.
Such regrouping can be done by specifying a threshold $\delta$ and to make each page
whose node parameter exceeds $\delta$ as a separate group (that is, a single group). 
This process can be carried out at the level of domains in a distributed way. 
For example, for $\delta=0.4$, the grouping and thus the node parameters 
changed as shown in Fig.~\ref{fig:delta_i_04}. Here, the number of groups has 
increased from 684 to 1,357 while the largest group has decreased
in size from 2,467 pages to 2,386 pages.
It should be noted that the removal of nodes from a group might also change 
the node parameters for pages that remain in the same group. Hence,
it is usually necessary to iterate the process several times
before all $\delta_i$ become below the given threshold. 

\begin{figure}[t]
  \centering
  \fig{9cm}{!}{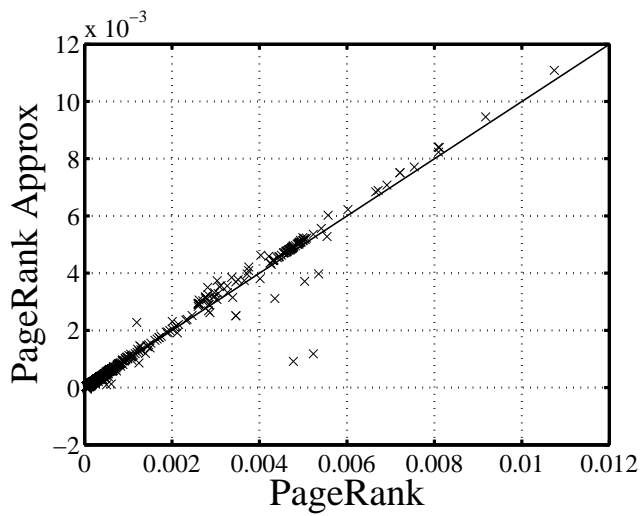}
  \caption{PageRank $x^*$ and its approximate $x'$ with the node parameter $\delta=0.4$.
      For most pages, the error in the approximation is small, resulting in a linear model
      via least-square fitting with slope 1.013.}
  \label{fig:PR_orig_approx}
\end{figure}

Once the grouping is settled, we compute the approximate value $x'$ of PageRank
via the aggregated approach. In the case with $\delta=0.4$, the error in the 
approximation seems small, where the total error was calculated to be
$\|x'-x^*\|_1=0.0665$.
More precisely, the relation for each page between the true PageRank $x^*$ and its
approximate $x'$ is shown in Fig.~\ref{fig:PR_orig_approx}. 
We made a linear model by least-square fitting, which 
resulted in the line with slope of 1.013 shown in the plot. 
Though this slope is very close to the desired 1.0, there are several points far from the line. 
The level of approximation can also be measured 
by computing the sample correlation between the two vectors $x^*$ and $x'$.
The Pearson correlation \cite{SneCoc:89}, representing the similarity in the values, is very high
at 0.991. On the other hand, the Spearman correlation
\cite{SneCoc:89} 
is related to the closeness
in the rankings among the pages and turned out to be 0.906. 
It should be noted that our implementation of (re-)grouping the pages has been performed 
under very simple rules, and there certainly is room for improvement. For example,
pages can be grouped by considering not only their domains but also 
their directories, subdirectories, related sites, etc.

\begin{figure}[t]
  \centering
  \fig{9cm}{!}{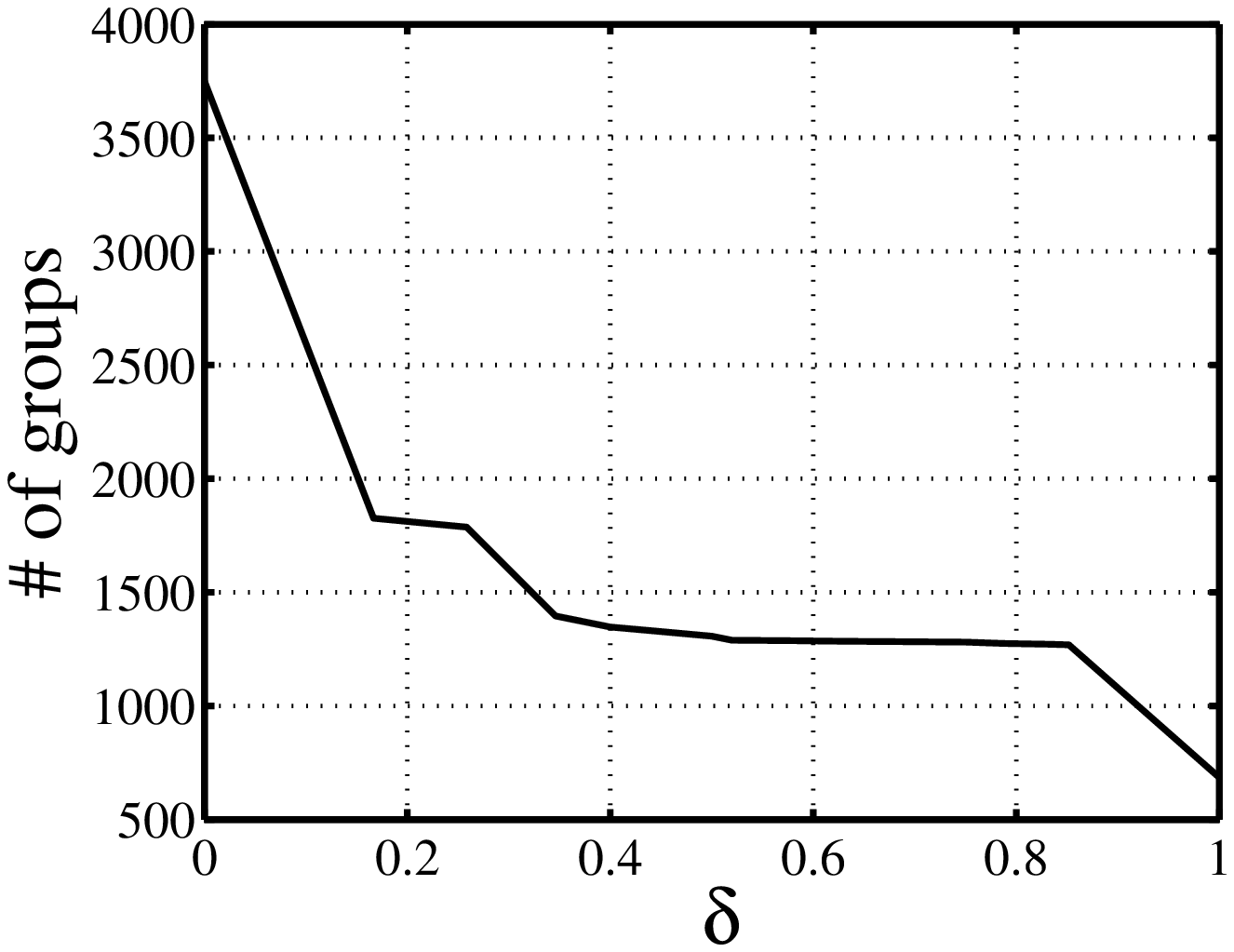}
  \caption{The number of groups versus the node parameter $\delta$. By reducing the size of $\delta$,
  the number of groups increases since any pages with parameter $\delta_i>\delta$ are taken out of 
   the group and then turned into single groups.}
  \label{fig:num_grp_delta}
\end{figure}

\begin{figure}[t]
  \centering
  \fig{9cm}{!}{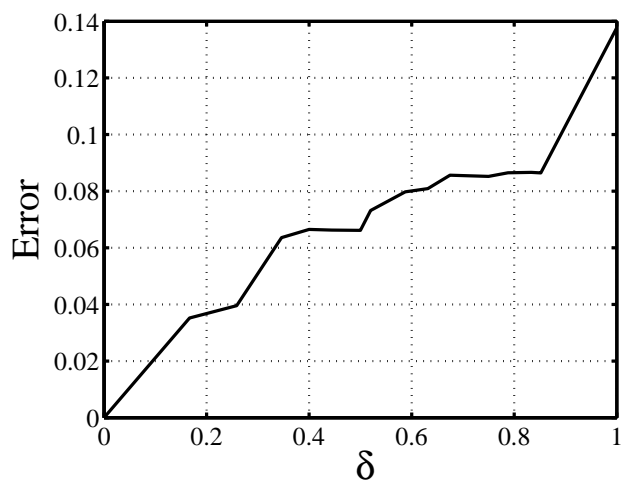}
  \caption{The error $\norm{x'-x^*}_1$ in approximated PageRank versus the node parameter $\delta$.
           Smaller $\delta$ results in better approximation.}
  \label{fig:error_delta}
\end{figure}

\begin{figure}[t]
  \centering
  \fig{9cm}{!}{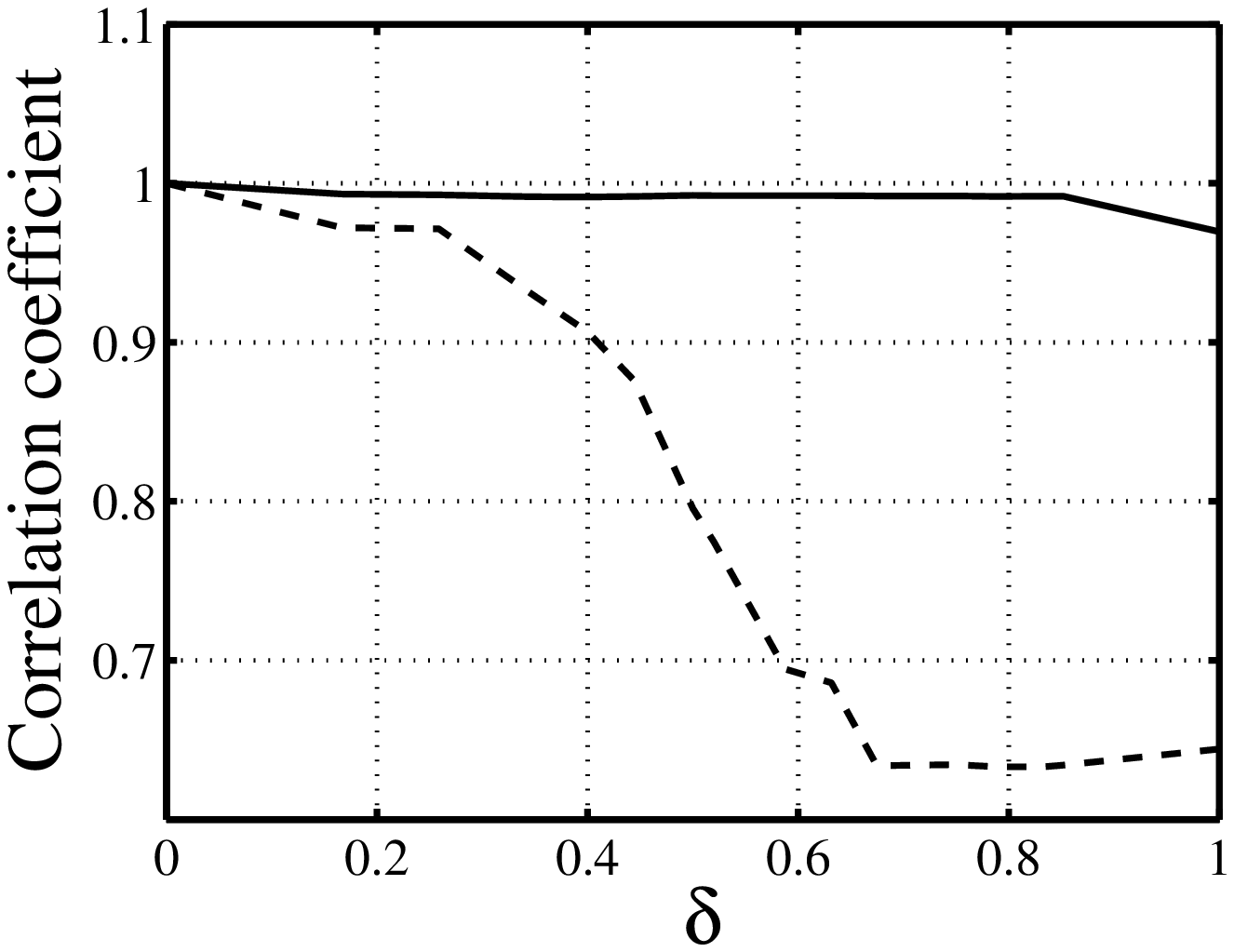}
  \caption{Pearson correlation (solid) and Spearman correlation (dashed) versus the
           node parameter $\delta$. The error in the approximated PageRank $x'$ can 
            be observed through these sample correlations between $x^*$ and $x'$.}
  \label{fig:correlation_delta}
\end{figure}

Similar computations can be made for different threshold values $\delta$. The results
are shown in Figs.~\ref{fig:num_grp_delta}--\ref{fig:correlation_delta},
which, respectively, display the number of groups, the error in the approximated 
PageRank, and the correlations versus the node parameter $\delta$.
Overall, the curves in these plots are smooth, showing that the grouping
method is sufficiently sensitive to changes in the threshold $\delta$.
In particular, it is interesting to observe that between $\delta=0.3$ and $0.8$, the
number of groups do not change much (Fig.~\ref{fig:num_grp_delta}),
but the reduction in approximation error (Fig.~\ref{fig:error_delta})
is very large when $\delta$ is reduced in size. Such a property is also demonstrated 
in the improvement in correlations (Fig.~\ref{fig:correlation_delta}).

Finally, computational aspects of the proposed algorithms are
briefly discussed.
In the approximated PageRank, the aggregated part can be computed through the
distributed algorithm similarly to the full-order case explained above. 
In the distributed randomized algorithm with $\delta=0.4$, the aggregated state 
is of order 1,347. 
In Fig.~\ref{fig:timeresp_err1}, the error is shown by the dashed line 
in comparison with the original distributed algorithm of full order (in the solid line).
Note that this error is obtained from the entire vector $x$ constructed at 
each time step and is with respect to the true PageRank
(and not the approximated version). Consequently,
the error stops decreasing after it reaches about 0.0665 as the vector $x$ converges 
to $x'$.
It is clear that the convergence speed is faster than that of the 
non-aggregated case in the solid line. 
This speed enhancement is in fact achieved with overall less computation; see \cite{IshTemBai:12}
for further discussion on computational costs.

\section*{Conclusion}
\label{sec:concl}

PageRank is a paradigmatic problem 
of great interest when ``big data" is available, and algorithms derived 
from PageRank have been successfully used to rank different objects in 
order of importance, such as scientific papers linked by citations, 
authors related by co-authorship, proteins in system biology and professional 
athletes. Therefore, in addition to systems and control, this problem is 
attracting the attention of many researchers working in a diverse set of fields, 
such as computer engineering, communications, physics, numerical analysis, 
linear algebra, and graph theory. 

The computation of PageRank is difficult due to the size of the web and because it is hard to gather and use global information about the network structure. 
In this article, we have followed a randomized decentralized approach, which leads to distributed
and parallel implementation, 
to deal with the extremely heavy computational load involved in the PageRank computation. The efficacy of the proposed approach
has been analyzed using the database \cite{webdata} 
(which has been previously used as a benchmark for PageRank algorithms \cite{FABG:13})
collected by crawling web pages of various universities.
To deal with problems of larger scale, the aggregation-based method
may repeatedly be applied in a hierarchical manner, by partitioning the initial groups,
then further the subgroups, and so on. 
Analyzing such a method is left for future research.

\newpage
\bibliographystyle{unsrt}


\end{document}